\pdfoutput=1

\documentclass[journal]{IEEEtran}
\usepackage{epsfig}
\usepackage{graphicx}
\usepackage{amsmath}
\usepackage{amssymb}
\usepackage{multirow}
\usepackage{cite}
\usepackage[table,xcdraw]{xcolor}
\usepackage[normalem]{ulem}

\useunder{\uline}{\ul}{}
\ifCLASSINFOpdf
\else
\fi
\begin{document}
%
% paper title
% Titles are generally capitalized except for words such as a, an, and, as,
% at, but, by, for, in, nor, of, on, or, the, to and up, which are usually
% not capitalized unless they are the first or last word of the title.
% Linebreaks \\ can be used within to get better formatting as desired.
% Do not put math or special symbols in the title.
\title{MDCN: Multi-scale Dense Cross Network for Image Super-Resolution}

\author{Juncheng Li, Faming Fang, Jiaqian Li, Kangfu Mei, and Guixu Zhang

\thanks{J. Li, F. Fang, J. Li, and G. Zhang are with the Shanghai Key Laboratory of Multidimensional Information Processing, East China Normal University, Shanghai 200062, China, and also with the School of Computer Science and Technology, East China Normal University, Shanghai 200062, China.
E-mail: cvjunchengli@gmail.com, \{fmfang, gxzhang\}@cs.ecnu.edu.cn}% <-this % stops a space
\thanks{K. Mei is with The Chinese University of Hong Kong, ShenZhen, China.}}% <-this % stops a space
%\thanks{Manuscript received April 19, 2005; revised August 26, 2015.}}

% The paper headers
\markboth{IEEE}
%\markboth{IEEE Transactions on Circuits and Systems for Video Technology}%
{Shell \MakeLowercase{\textit{et al.}}: Bare Demo of IEEEtran.cls for IEEE Journals}
% The only time the second header will appear is for the odd numbered pages
% after the title page when using the twoside option.
%
% *** Note that you probably will NOT want to include the author's ***
% *** name in the headers of peer review papers.                   ***
% You can use \ifCLASSOPTIONpeerreview for conditional compilation here if
% you desire.

% If you want to put a publisher's ID mark on the page you can do it like
% this:
%\IEEEpubid{0000--0000/00\$00.00~\copyright~2015 IEEE}
% Remember, if you use this you must call \IEEEpubidadjcol in the second
% column for its text to clear the IEEEpubid mark.

% use for special paper notices
%\IEEEspecialpapernotice{(Invited Paper)}

% make the title area
\maketitle

% As a general rule, do not put math, special symbols or citations
% in the abstract or keywords.
\begin{abstract}
Convolutional neural networks have been proven to be of great benefit for single-image super-resolution (SISR).
However, previous works do not make full use of multi-scale features and ignore the inter-scale correlation between different upsampling factors, resulting in sub-optimal performance.
Instead of blindly increasing the depth of the network, we are committed to mining image features and learning the inter-scale correlation between different upsampling factors.
To achieve this, we propose a Multi-scale Dense Cross Network (MDCN), which achieves great performance with fewer parameters and less execution time.
MDCN consists of multi-scale dense cross blocks (MDCBs), hierarchical feature distillation block (HFDB), and dynamic reconstruction block (DRB).
Among them, MDCB aims to detect multi-scale features and maximize the use of image features flow at different scales, HFDB focuses on adaptively recalibrate channel-wise feature responses to achieve feature distillation, and DRB attempts to reconstruct SR images with different upsampling factors in a single model.
It is worth noting that all these modules can run independently.
It means that these modules can be selectively plugged into any CNN model to improve model performance.
Extensive experiments show that MDCN achieves competitive results in SISR, especially in the reconstruction task with multiple upsampling factors.
The code will be provided at https://github.com/MIVRC/MDCN-PyTorch.
\end{abstract}

% Note that keywords are not normally used for peerreview papers.

% For peer review papers, you can put extra information on the cover
% page as needed:
% \ifCLASSOPTIONpeerreview
% \begin{center} \bfseries EDICS Category: 3-BBND \end{center}
% \fi
%
% For peerreview papers, this IEEEtran command inserts a page break and
% creates the second title. It will be ignored for other modes.
\IEEEpeerreviewmaketitle

\section{Introduction}\label{Introduction}
% The very first letter is a 2 line initial drop letter followed
% by the rest of the first word in caps.
%
% form to use if the first word consists of a single letter:
% \IEEEPARstart{A}{demo} file is ....
%
% form to use if you need the single drop letter followed by
% normal text (unknown if ever used by the IEEE):
% \IEEEPARstart{A}{}demo file is ....
%
% Some journals put the first two words in caps:
% \IEEEPARstart{T}{his demo} file is ....
%
% Here we have the typical use of a "T" for an initial drop letter
% and "HIS" in caps to complete the first word.

\IEEEPARstart{I}{mage} super-resolution, especially single image super-resolution (SISR) is an extremely hot topic in the computer vision field, which aims to reconstruct a super-resolution (SR) image from its degraded low-resolution (LR) one.
In order to generate high-quality SR images, plenty of SR methods have been proposed, including interpolation-based~\cite{zhang2006edge, liu2011image, dong2013sparse, zhang2018single}, anchored neighborhood regression based~\cite{timofte2013anchored, timofte2014a+}, self-example learning based~\cite{yang2012self, zhu2014fast, huang2015single}, and learning based~\cite{dong2014learning, kim2016accurate} methods.

Recently, convolutional neural networks (CNNs) have achieved great success in computer vision tasks, which also profoundly promote the development of SISR.
Therefore, CNN-based SR methods~\cite{dong2014learning, dong2016accelerating, kim2016accurate, kim2016deeply, shi2016real, lai2017deep, lim2017enhanced, tai2017image, tong2017image, haris2018deep, han2018image, zhang2018learning, He2019CascadedDN, wu2020multi, zuo2019multi, Hu2018ChannelwiseAS, Li2019FilterNetAI, he2019mrfn, li2019lightweight, Wang2019ResolutionAwareNF, Xie2019FastSS, fang2020soft} have become the mainstream today, which aim to learn the mapping between LR and HR images by constructing a well-designed network.
Among them, Dong et al.~\cite{dong2014learning} proposed the SRCNN, which was the first successful model adopting CNN to the SISR task.
After that, Kim et al.~\cite{kim2016accurate} introduced residual learning into SISR and proposed the VDSR;
Lim et al.~\cite{lim2017enhanced} introduced residual block and multi-factor strategy to construct EDSR and MDSR models.
Li et al.~\cite{Li_2018_ECCV} proposed the MSRN by introducing multi-scale residual learning;
Zhang et al.~\cite{Zhang_2018_ECCV} introduced the attention mechanism into residual blocks to construct a very deep RCAN;
He et al.~\cite{He2019CascadedDN} presented a cascaded network (CDN\_MRF) with multi-receptive fields to increase the spatial resolution; 
Li et al.~\cite{Li2019FilterNetAI} proposed a deep adaptive information filtering network (FilterNet) for accurate and fast SR image reconstruction;
Apart from these models, more CNN-based SR models can be found in~\cite{yang2019deep} and~\cite{Wang2020DeepLF}.

\begin{figure}
   \begin{center}
   \includegraphics[width=0.9\linewidth]{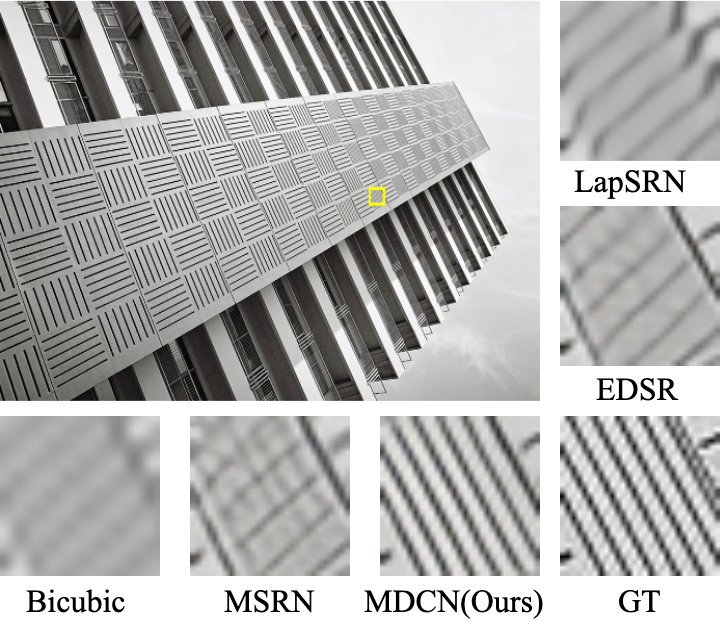}
   \end{center}
   \caption{Visual comparisons of $\times$3 SR on a challenging image.
   The SR images reconstructed by other models have been severely blurred, but our MDCN can suppress artifacts and reconstruct clear texture details.}
   \label{Head}
\end{figure}

Although SR models mentioned above can achieve great results (Fig.~\ref{Head}), they are usually accompanied by complicated structures and huge computational overhead.
Among all of them, MSRN~\cite{Li_2018_ECCV} is the focus of this paper.
MSRN~\cite{Li_2018_ECCV} was proposed in 2018, which used multi-scale features to reconstruct SR images.
To achieve this, the multi-scale residual block (MSRB) was proposed for feature extraction.
Now, many works~\cite{cao2019single,sang2019multi,qin2020multi,li2019multi,soh2020lightweight,wang2020remote,zhang2020scene,zhang2019multi,qi2020pulmonary,sang2019contourlet} have proved the effectiveness of multi-scale residual learning strategy and multi-scale features have been used in various fields, such as image denoising~\cite{sang2019contourlet}, remote sensing image super-resolution~\cite{wang2020remote,zhang2020scene}, and medical image enhancement~\cite{zhang2019multi,qi2020pulmonary}.
These models directly migrate MSRN to other tasks or by optimizing the model structure of MSRN to improve model performance.
Though these models greatly improved the architecture and application of MSRN, none of them solved the core problem of MSRN.
In this work, we aim to propose a more efficient and general model.

\textbf{(A). Multi-scale Feature Extraction:} Massive studies~\cite{szegedy2015going, szegedy2016rethinking, chollet2017xception, lai2017fast} have pointed out that image will exhibit different characteristics at different scales and making full use of these features can further improve model performance.
Therefore, Li et al.~\cite{Li_2018_ECCV} proposed a MSRB for multi-scale feature extraction, which integrated different convolutional kernels in a block to adaptively extract image features at different scales.
However, it does not make full use of image features in the previous layers, so local features are difficult to transfer to other layers and the multi-scale feature flow will be suppressed.
Therefore, exploring a more efficient multi-scale feature extraction block is essential for image reconstruction.

\textbf{(B). Hierarchical Feature Distillation:} As the depth of the network increases, image features will gradually disappear during the conduction process.
Therefore, taking advantage of hierarchical features will greatly improve model performance.
However, most models ignore this or simply concatenate all hierarchical features.
These methods cannot eliminate redundant features, resulting in sub-optimal results and inefficient image reconstruction.
Therefore, an effective method that can exploit hierarchical features and eliminate redundant features is crucial for SISR.

\textbf{(C). Inter-scale Correlation Exploration:}
Most SR models introduce the deconvolutional layer or sub-pixel convolutional layer to achieve image magnification.
However, due to their characteristics, these models need to train specific models for different upsampling factors.
Although some models introduce simple and flexible reconstruction modules, they still fail to achieve the expectation that a model can be suitable for multiple upsampling factors.
Therefore, a model that can adapt to multiple factors and learn the inter-scale correlation between different upsampling factors is needed.

To solve these issues, we propose a Multi-scale Dense Cross Network (MDCN).
MDCN consists of multi-scale dense cross blocks (MDCBs), hierarchical feature distillation block (HFDB), and dynamic reconstruction block (DRB).
Among them, MDCB is an efficient feature extraction module that can extract rich high-frequency details through the integrated dual-path dense network and multi-scale learning.
HFDB introduces dimension transformation and channel attention mechanism to adaptively recalibrate channel-wise feature responses, thus redundant hierarchical features will be removed.
DRB aims to maximize the reuse of model parameters and learn the inter-scale correlation between different upsampling factors by dynamic activating the corresponding upsampling module.
In summary, our contributions are as follows:

\begin{table}[t]
   \scriptsize
   \centering
   \setlength{\tabcolsep}{3.5mm}
   \renewcommand\arraystretch{1}
   \caption{Comparisons of different feature extraction blocks.
   $\surd$ and $\times$ indicate whether the block uses the corresponding mechanism.}
   \begin{tabular}{l|c|c|c}
   \hline
   Block  & \begin{tabular}[c]{@{}c@{}}Residual \\ Learning\end{tabular} & \begin{tabular}[c]{@{}c@{}}Dense \\ Connection\end{tabular} & \begin{tabular}[c]{@{}c@{}}Multi-scale \\ Learning \end{tabular}\\ \hline\hline
   Residual Block ~\cite{lim2017enhanced}      & $\surd$           & $\times$         & $\times$     \\
   Dense Block ~\cite{tong2017image}      & $\times$          & $\surd$          & $\times$    \\
   RDB ~\cite{zhang2018residual}         & $\surd$           & $\surd$          & $\times$    \\
   MSRB ~\cite{Li_2018_ECCV}             & $\surd$           & $\times$         & $\surd$     \\
   \textbf{MDCB (Ours)}        & $\surd$           & $\surd$          & $\surd$      \\
   \hline
   \end{tabular}
   \label{Feature Block}
\end{table}

(i) We propose a Multi-scale Dense Cross Network (MDCN) for SISR, which achieves competitive results with fewer parameters and less execution time.

(ii) We devise a Multi-scale Dense Cross Block (MDCB) for feature extraction, which is essentially a dual-path dense network that effectively detects local and multi-scale features.

(iii) We design a Hierarchical Feature Distillation Block (HFDB) to maximize the use of hierarchical features. 
It is the first CNN module specifically designed for hierarchical feature learning.

(iv). We introduce the Dynamic Reconstruction Block (DRB) to learn the inter-scale correlation between different upsampling factors, which enables MDCN to reconstruct SR images with multiple factors in a single model.

\section{Related Work}\label{hierarchical}
SISR has attracted increasing attention in recent years since the quality of reconstructed SR images will seriously affect the accuracy of high-level computer vision tasks such as image classification~\cite{Russell2019FeatureBasedIP, Han2020DoubleRR}, image segmentation~\cite{Wang2019HierarchicalIS, Zhou2018ComputationAM}, and object detection~\cite{Li2020HeadNetAE, Sommer2019ComprehensiveAO}.
In all relevant studies, feature extraction and multi-factor model have attracted our attention.

\begin{figure}
   \begin{center}
   \includegraphics[width=0.8\linewidth]{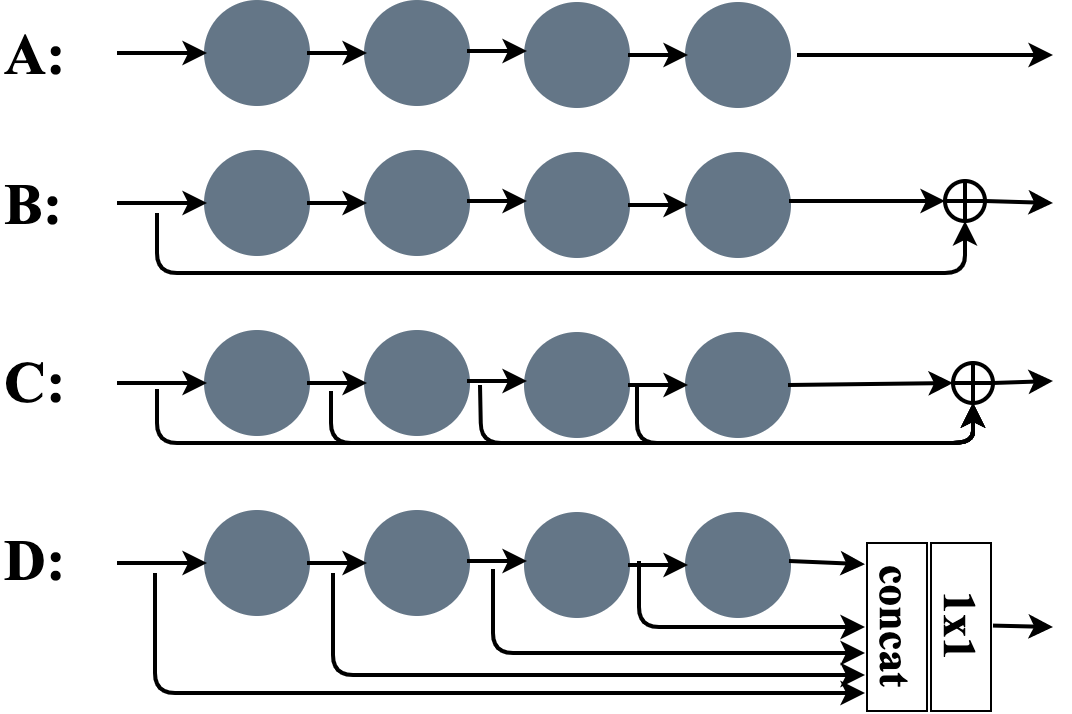}
   \end{center}
   \caption{Comparisons of different hierarchical feature utilization methods.}
   \label{Connection}
\end{figure}

\textbf{Feature Extraction}:
Recently, numerous feature extraction blocks have been proposed for local feature extraction.
In TABLE~\ref{Feature Block}, we provide several classic feature extraction blocks, including Residual block~\cite{he2016deep}, Dense block~\cite{huang2017densely}, Residual Dense Block (RDB~\cite{zhang2018residual}), and Multi-scale Residual Block (MSRB~\cite{Li_2018_ECCV}).
According to the table, we can clearly see the mechanism introduced in each block.
Different from previous feature extraction blocks, MSRB introduces a new multi-scale learning strategy.
Although this allows MSRB to obtain multi-scale features, it ignores the use of features in the previous layers.
In this paper, by rethinking the influence of residual learning, dense connection, and multi-scale learning, we aim to explore a new feature extraction block that can skillfully combine these strategies without adding additional parameters.

On the other hand, as the depth of the network increases, image features will be gradually lost in the process of transmission.
Therefore, making full use of hierarchical features is also important for image restoration.
In Fig.~\ref{Connection}, we show four different methods, including the series connection (A, without using hierarchical feature), residual connection (B, only use the input image features), dense residual connection (C, use all hierarchical features for residual learning), and hierarchical connection (D, concatenate all hierarchical features and apply a $1\times1$ layer for feature fusion).
However, by investigating the performance of methods A-D, we find that hierarchical features can improve the model performance while redundant hierarchical features will make the model more difficult to train.
Therefore, a new method that can effectively utilize hierarchical features and eliminate redundant features is crucial.

\begin{figure}
   \begin{center}
   \includegraphics[width=0.85\linewidth]{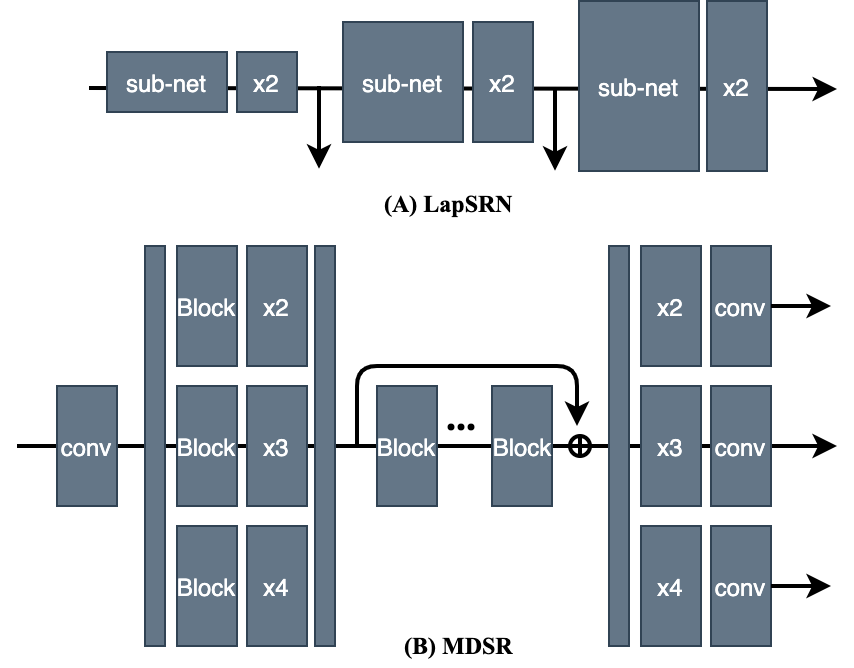}
   \end{center}
   \caption{The main structure of LapSRN~\cite{lai2017deep} and MDSR~\cite{lim2017enhanced}.}
   \label{MUF}
\end{figure}

\textbf{Multi-factor Model}:
A model that can be applied to multiple upsampling factors is the hot topic in SISR, which is the basic condition for inter-scale correlation learning.
Nonetheless, due to the limitations of deconvolutional layer and sub-pixel convolutional layer, there is still no perfect solution for this problem.
Recently, some models have been proposed for multiple upsampling factors, such as LapSRN~\cite{lai2017deep} and MDSR~\cite{lim2017enhanced}.
LapSRN~\cite{lai2017deep} (Fig.~\ref{MUF} (A)) is a Laplacian pyramid framework, which can progressively reconstruct high-resolution images.
However, this method essentially decomposes the large upsampling factor into multiple small upsampling factors for image reconstruction, which does not consider the inter-scale correlation between different upsampling factors.
MDSR~\cite{lim2017enhanced} (Fig.~\ref{MUF} (B)) introduces the scale-specific processing modules at the head and tail of the model to handle different upsampling factors.
This structure can be suitable for multiple upsampling factors, but the scale-specific processing modules at the head of the model are not conducive to the transmission of information and the interaction of inter-scale features.
In order to maximize the reuse of model parameters and further exploit the inter-scale correlation between different upsampling factors, we aim to further optimize the strategy used in MDSR and propose a more efficient method.

Notice that some papers term the models that can be used for multiple upsampling factors as 'multi-scale network'.
In this paper, we name this type of model as 'multi-factor model' to distinguish it from the multi-scale feature extraction block.

\section{Multi-scale Dense Cross Network (MDCN)}
\begin{figure*}
   \begin{center}
   \includegraphics[width=0.85\linewidth]{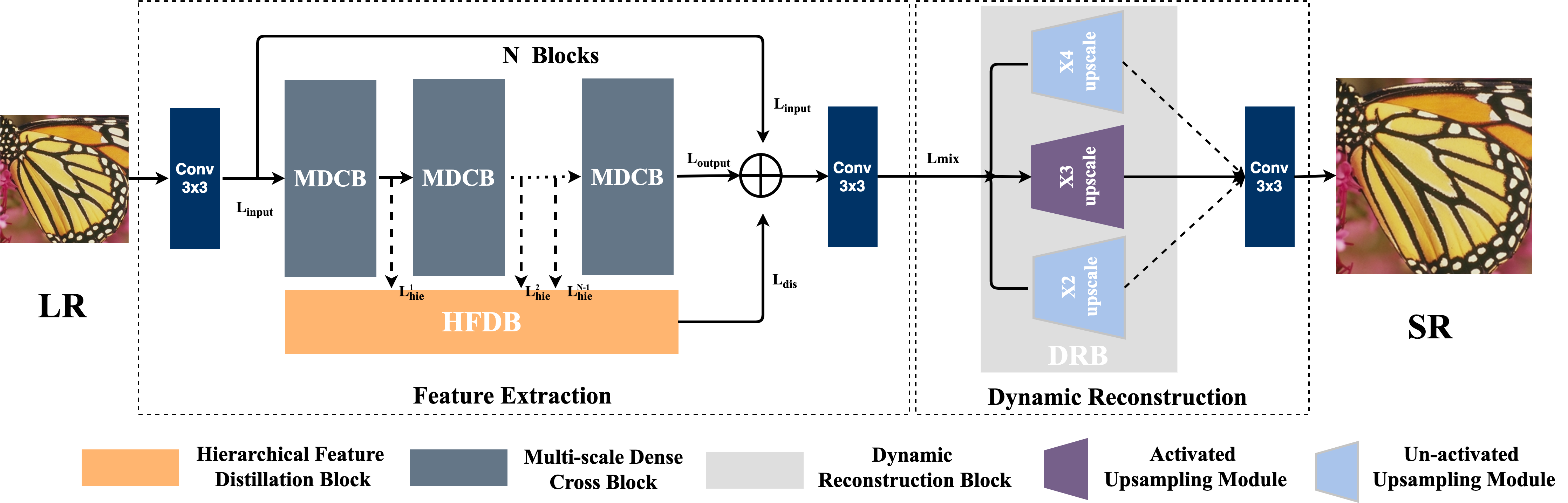}
   \end{center}
   \caption{The complete architecture of our proposed Multi-scale Dense Cross Network (MDCN), which consists of two stages: feature extraction and dynamic reconstruction. The dark grayish block, orange block, and gray block denote the MDCB, HFDB, and DRB, respectively.}
   \label{MDCN}
\end{figure*}

In this paper, we propose a Multi-scale Dense Cross Network (MDCN) for SISR.
As shown in Fig.~\ref{MDCN}, MDCN can be divided into two stages: feature extraction and dynamic reconstruction.
In stage I, we use $N$ multi-scale dense cross blocks (MDCBs) and a hierarchical feature distillation block (HFDB) for local feature extraction and hierarchical features distillation, respectively.
In stage II, we introduce a dynamic reconstruction block (DRB) for SR image reconstruction.
This block allows the model can be suitable for multiple upsampling factors, which makes MDCN more scalable.

Define $I_{\rm LR}$ and $I_{\rm SR}$ as the input and output of MDCN.
$L_{\rm input}$, $L_{\rm output}$, and $L_{\rm mix}$ are the input of the first MDCB, the output of the last MDCB, and the input of DRB, respectively.
Following previous works, we first use a 3$\times$3 convolutional layer to upgrade the LR image to a high dimensional
\begin{equation}
   \small
   L_{\rm input} = F_{\rm input}(I_{\rm LR}),
\end{equation}
where $F_{\rm input}(\cdot)$ denotes the corresponding convolutional layer and $L_{\rm input}$ is the converted features. 
Meanwhile, $L_{input}$ is also served as the input of MDCB for local feature extraction
\begin{equation}
   \small
   L_{\rm output} = F_{\rm MDCG}(L_{\rm input}),
\end{equation}
where $F_{\rm MDCG}(\cdot)$ represents the multi-scale dense cross group (MDCG), which consists of $N$ MDCBs.
To make full use of the hierarchical features, we also introduce the HFDB for hierarchical feature fusion and distillation
\begin{equation}
   \small
   L_{\rm dis} = F_{\rm HFDB}([L_{\rm hie}^{1}, L_{\rm hie}^{2},...,L_{\rm hie}^{N-1}]), 
\end{equation}
where $F_{\rm HFDB}(\cdot)$ denotes the HFDB and $L_{\rm hie}^{n}$ ($n=1,2,...,N-1$) represents the output of the $n$-th MDCB. 
In addition, $L_{\rm dis}$ denotes the distilled hierarchical features, which can be used for image reconstruction.

After feature extraction, we combine the original image feature $L_{\rm input}$, the extracted high-level features $L_{\rm output}$, and the distilled hierarchical features $L_{\rm dis}$ for the final SR image reconstruction
\begin{equation}
   \small
   L_{\rm mix} = F_{\rm mix}(L_{\rm input} + L_{\rm dis} + L_{\rm output}),
\end{equation}
where $F_{\rm mix}(\cdot)$ is a 3$\times$3 convolutional layer used for feature fusion.
Finally, the merged features are delivered to the DRB for high-quality SR image reconstruction and a 3$\times$3 convolutional layer is applied
to convert it to RGB space
\begin{equation}
   \small
   I_{\rm SR} = F_{\rm output}(F_{\rm DRB}(L_{\rm mix})),
\end{equation}
where $F_{\rm DRB}(\cdot)$ denotes the dynamic reconstruction block, $F_{\rm output}(\cdot)$ represents the corresponding operation that converts image feature maps from high dimensional space to RGB space, and $I_{\rm SR}$ is the finally reconstructed SR image.

During training, MDCN is optimized with $\rm L1$ loss function.
Therefore, Given a training dataset $\left \{I_{\rm LR}^{i}, I_{\rm HR}^{i} \right \}_{i=1}^{M}$, we solve
\begin{equation}
   \small
   \hat{\theta} = arg\,\min\limits_{\theta}\, \frac{1}{M}\sum_{i=1}^{M} \left \| F_{\theta}(I_{\rm LR}^{i}) - I_{\rm HR}^{i} \right \|_{1},
\end{equation}
where $\theta$ denotes the parameter set of our model and $F(\cdot)$ denotes the MDCN.
Each module of MDCN will be described in the following section.

\subsection{Multi-scale Dense Cross Block (MDCB)} \label{Method-MDCB}
Features extraction is the most important step in image restoration.
In order to make full use of image features from the LR image, we propose a new module named multi-scale dense cross block (MDCB).
MDCB is the basic component of MDCN, which is inspired by MSRB~\cite{Li_2018_ECCV} and aims to solve the problem that MSRN can not obtain features from the previous layers.
Compared with MSRB~\cite{Li_2018_ECCV}, MDCB adopts a dual dense network as the backbone, which enables it to extract image features at different scales as well as take advantage of features from the previous layers.

\begin{figure}
   \begin{center}
   \includegraphics[width=0.72\linewidth]{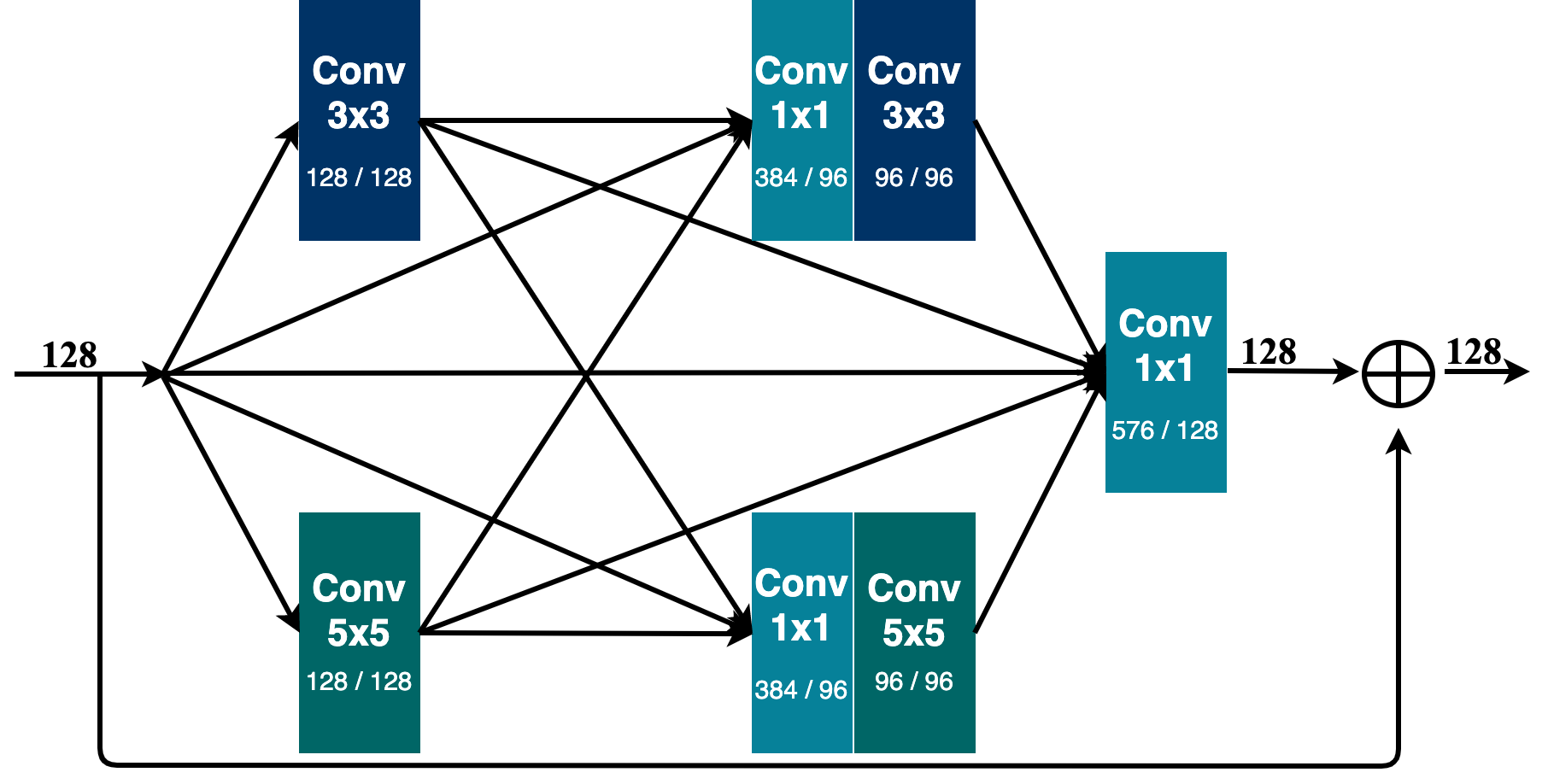}
   \end{center}
   \caption{The architecture of multi-scale dense cross block (MDCB).}
   \label{MDCB}
\end{figure}

\begin{figure*}
   \begin{center}
   \includegraphics[width=0.9\linewidth]{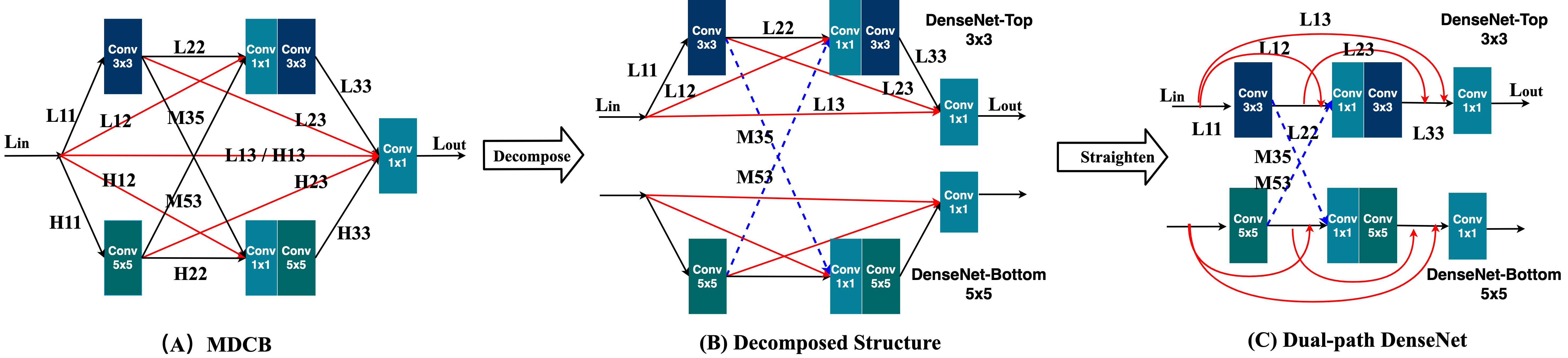}
   \end{center}
   \caption{The decomposed structure of MDCB, which remove the residual learning for better representation.
   (A) is the MDCB structure that removes residual learning, (B) is the decomposed structure of MDCB, and (C) is the equivalent structure after straightening (B).}
   \label{MDCB-D}
\end{figure*}

\textbf{Dual-path Network:}
As shown in Fig.~\ref{MDCB}, MDCB is essentially a dual-path network, which uses two dense networks as the backbone.
In order to better explain the working mechanism of the module, we provide the decomposed structure of MDCB in Fig.~\ref{MDCB-D}.
In Fig.~\ref{MDCB-D}, (A) is the simplified version of MDCB, which removes residual learning for better explanation, (B) is the decomposed structure of MDCB, and (C) is the equivalent structure after straightening (B).
As shown in Fig.~\ref{MDCB-D} (C), the model can be divided into two parts: DenseNet-Top and DenseNet-Bottom.
Each part is actually a modified dense network~\cite{huang2017densely}, which reduces the depth of the network and introduces bottleneck layers for features fusion.
Following the dense network, our Dense-Top or DenseNet-Bottom connects each layer to every other layers in a feed-forward fashion (red lines), which can detect rich features in the previous layers.
The operations of Dense-Top are:
\begin{equation}
   \small
   L_{\rm 22} = C^{1}_{\rm 3\times3}(L_{\rm 11}),
\end{equation}
\begin{equation}
   \small
   L_{\rm 33} = C^{2}_{\rm 3\times3}(C^{2}_{\rm 1\times1}([L_{\rm 12}, L_{\rm 22}])),
\end{equation}
\begin{equation}
   \small
   L_{\rm out} = C^{3}_{\rm 1\times1}([L_{\rm 13}, L_{\rm 23}, L_{\rm 33}]),
\end{equation}
where $C^{p}_{\rm 3\times3}$ denotes the $p$-depth 3$\times$3 convolutional layer, $C^{p}_{\rm 1\times1}$ denotes the $p$-depth 1$\times$1 convolutional layer, and $[\cdot,\cdot,\cdot]$ denotes the concatenation operation.
In this part, we use 3$\times$3 convolutional layers for feature extraction and use 1$\times$1 convolutional layer for feature fusion.
Inspired by MSRB~\cite{Li_2018_ECCV}, we also introduce multi-scale learning to extract image features at different scales.
Therefore, we replace all 3$\times$3 convolutional layers with 5$\times$5 convolutional layers in the DenseNet-Bottom
\begin{equation}
   \small
   H_{\rm 22} = C^{1}_{\rm 5\times5}(H_{\rm 11}),
\end{equation}
\begin{equation}
   \small
   H_{\rm 33} = C^{2}_{\rm 5\times5}(C^{2}_{\rm 1\times1}([H_{\rm 12}, H_{\rm 22}])),
\end{equation}
\begin{equation}
   \small
   H_{\rm out} = C^{3}_{\rm 1\times1}([H_{\rm 13}, H_{\rm 23}, H_{\rm 33}]).
\end{equation}

\textbf{Feature Exchange \& Fusion Mechanism:}
As mentioned above, MDCB uses two dense networks to extract image features at different scales.
However, if these two subnets are independent of each other, image features at different scales will be difficult to transfer and fuse.
Therefore, we associate the DenseNet-Top and DenseNet-Bottom by introducing two skip connections: $M_{\rm 35}$ and $M_{\rm 53}$.
As shown in Fig.~\ref{MDCB-D}, the blue lines are the introduced skip connection.
$M_{\rm 35}$ and $M_{\rm 53}$ transmit the unique features extracted by themselves to the other sub-net.
This facilitates the exchange of image features at different scales and improves the model performance.
Meanwhile, we also introduce a bottleneck layer at the tail of the module to achieve feature fusion and dimensionality reduction, which combines the multi-scale features ($L_{\rm 33}$ and $H_{\rm 33}$) extracted by DenseNet-Top and DenseNet-Bottom with the local features ($L_{\rm 23}$, $H_{\rm 23}$, and $L_{\rm 13}/H_{\rm 13}$) provided by the previous layers to obtain the high-frequency features.
Therefore, the complete operations of multi-scale dense feature extraction can be defined as
\begin{equation}
   \small
   L_{\rm 22} = C^{1}_{\rm 3\times3}(L_{\rm 11}), H_{\rm 22} = C^{1}_{\rm 5\times5}(H_{\rm 11}),
\end{equation}
\begin{equation}
   \small
   L_{\rm 33} = C^{2}_{\rm 3\times3}(C^{2}_{\rm 1\times1}([L_{\rm 12}, L_{\rm 22}, M_{\rm 53}])),
\end{equation}
\begin{equation}
   \small
   H_{\rm 33} = C^{2}_{\rm 5\times5}(C^{2}_{\rm 1\times1}([H_{\rm 12}, H_{\rm 22}, M_{\rm 35}])),
\end{equation}
\begin{equation}
   \small
   L_{\rm out} = C^{3}_{\rm 1\times1}([L_{\rm 23}, L_{\rm 33}, H_{\rm 23}, H_{\rm 33}, L_{\rm in}]).
\end{equation}
Among them, $L_{\rm in}=L_{\rm 11}=L_{\rm 12}=L_{\rm 13}=H_{\rm 11}=H_{\rm 12}=H_{\rm 13}$, $L_{\rm 22}=L_{\rm 23}=M_{\rm 35}$, and $H_{\rm 22}=H_{\rm 23}=M_{\rm 53}$, which represent the same features while delivering to different convolutional layers.
In addition, $L_{\rm in}$ and $L_{\rm out}$ denote the input and output of MDCB, respectively.

\textbf{Local residual learning:}
Following previous works~\cite{Li_2018_ECCV, zhang2018residual}, we also introduce local residual learning into our MDCB to further improve the information flow.
Therefore, the output of $n$-th MDCB can be written as
$L_{n} = L_{n-1} + L_{\rm out}$.
$L_{n-1}$ denotes the output of the previous MDCB and serves as the input of this MDCB ($L_{\rm in} = L_{n-1}$).

\subsection{Hierarchical Feature Distillation Block (HFDB)}
\begin{figure}
   \begin{center}
   \includegraphics[width=0.9\linewidth]{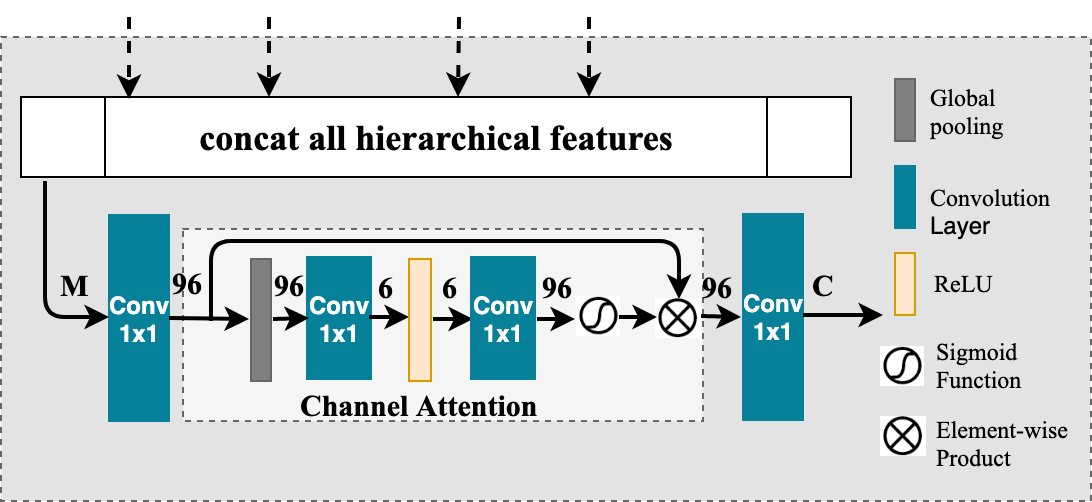}
   \end{center}
   \caption{The architecture of hierarchical feature distillation block (HFDB).}
   \label{Distillation}
\end{figure}

As mentioned in MSRN~\cite{Li_2018_ECCV}, hierarchical features can further improve model performance.
However, the method used in MSRN is crude.
This is because only use the 1$\times$1 convolutional layer to compress hierarchical features can not extract effective features.
Meanwhile, this method will produce massive redundant features, which is not conducive to model training.
In order to remove redundant features and fully mine useful hierarchical features, we propose a hierarchical feature distillation block (HFDB, Fig.~\ref{Distillation}).
HFDB focuses on adaptively recalibrate channel-wise feature response to achieve feature distillation, which is an efficient module that only needs low computational overhead.
The core of HFDB is the introduced dimension transformation and channel attention mechanism.

\textbf{Dimension Transformation Mechanism (DTM):}
Inspired by AutoEncoder, we introduce the dimension transformation mechanism to the module to achieve feature distillation.
As shown in Fig.~\ref{Distillation}, HFDB introduces the bottleneck layer (1$\times$1 layer) at the head and tail of the module, respectively.
Firstly, all hierarchical features are concatenate like MSRN~\cite{Li_2018_ECCV}.
Then, these $\rm M$ feature maps are sent to the bottleneck layer to reduce the feature dimension.
Then, the compressed features are sent to the channel attention module to adaptively recalibrate channel-wise feature response.
Finally, the dimension of recalibrated features will be upgraded by the bottleneck layer at the tail of the module.
It is worth noting that we set $\rm M > C >96$.
Therefore, this module achieves a similar effect to the AutoEncoder, thereby achieving feature distillation.

\textbf{Channel Attention Mechanism (CAM):}
As mentioned above, we introduce CAM to the module to mine the most useful hierarchical features.
CAM was firstly proposed in SENet~\cite{hu2018squeeze}, which could adaptively calibrate channel-wise feature response by explicitly modeling interdependency between channels.
Inspired by this, RCAN~\cite{Zhang_2018_ECCV} and SrSENet~\cite{mei2018effective} introduced CAM into all feature extraction blocks to obtain useful local features.
However, this method will only bring a slight performance improvement while the increasement in execution time and memory consumption is huge~\cite{li2019lightweight}. 
Different from RCAN~\cite{Zhang_2018_ECCV} and SrSENet~\cite{mei2018effective} that pay attention to the local features extraction, we focus on using CAM to mine the relationship between hierarchical features and extract the most useful hierarchical features.

Taking $X=[x_{1},...,x_{c},...,x_{C}]$ as input, which has $C$ feature maps with size of $H \times W$.
The $c$-th element of $z$ can be defined as
\begin{equation}
   \small
   z_{c} = \frac{1}{H \times W}\sum^{H}_{i=1}\sum^{W}_{j=1}x_{c}(i,j),
\end{equation}
where $x_{c}(i,j)$ is the value at position $(i,j)$ of $c$-th feature $x_{c}$.
Then, we apply a gating mechanism with the sigmoid function
$s = f(C_{1}(\delta (C_{2}(z)))),$
where $f(\cdot)$ and $\delta(\cdot)$ denote the sigmoid function and ReLU function, respectively.
$C_{1}(\cdot)$ is a convolutional layer, which acts as channel-downscaling with reduction ratio $r$ and is activated by ReLU, the low-dimensional feature is increased with ratio $r$ by a channel-upscaling layer $C_{2}(\cdot)$.
Finally, the channel statistics $s$ is used to rescale the input:
$\hat{x_{c}} = s_{c} \cdot x_{c},$
where $s_{c}$ and $x_{c}$ are the scaling factor and feature maps of the $c$-th channel.

Due to the ability of DTM and CAM, HFDB can not only eliminate redundant features but also extract useful hierarchical features.
Meanwhile, MDCN only uses the channel attention mechanism once, which can greatly improve the efficiency of the model.

\subsection{Dynamic Reconstruction Block (DRB)}
In order to exploit the inter-scale correlation between different upsampling factors and maximize the use of model parameters, we introduce the dynamic reconstruction block (DRB).
DRB is essentially a multi-element module that combines several different upsampling modules, which was first proposed in MDSR~\cite{lim2017enhanced}.
For single upsampling factor, DRB will create a corresponding scale-specific upsampling module.
However, as for multi-factor image super-resolution task, DRB will generate a set of corresponding scale-specific upsampling modules (e.g., $\times$2, $\times$3, and $\times$4).
In the upsampling module, we use the sub-pixel convolutional layer~\cite{shi2016real} to learn an array of upscaling filters to upscale feature maps into SR output.

As shown in Fig.~\ref{MDCN}, DRB is located at the tail of the model, which contains three sub-modules ($\times$2, $\times$3, and $\times$4 upsample modules).
Meanwhile, these upsampling modules are arranged in parallel and only one upsampling module will be activated when used. 
In other words, the data stream will only flow into the module corresponding to the upsampling factor for image reconstruction.
To better exploit the inter-scale correlation between different upsampling factors, we construct a dataset with different factors and introduce a mixed training strategy for training. 
This means that different size of images will be randomly sent to the model. 
Then it  will automatically activate the corresponding upsampling modules according to the upsampling factors. 
Compared with specifically training a model for each upsampling factor, this method can learn the latent inter-scale information, maximize the reuse of model parameters, and reduce the training time.
Thanks to this strategy, the trained model can be directly applied to multiple upsampling factors without re-training.

Different from MDSR~\cite{lim2017enhanced}, all parameters of our model are shared under different upsampling factors, except for the DRB.
This can maximize the reuse of model parameters and benefit for the model to learn the inter-scale correlation between different upsampling factors, thus improve the model performance and robustness.
In Sec.~\ref{SDRB}, we will further verify the effectiveness of this module.

\section{Experiments}\label{Exp}
Following previous works, we only use DIV2K (1-800)~\cite{agustsson2017ntire} as our training dataset.
Meanwhile, we choose Set5~\cite{bevilacqua2012}, Set14~\cite{zeyde2010single}, B100~\cite{martin2001database}, Urban100~\cite{huang2015single}, and Manga109~\cite{matsui2017sketch} as our test datasets.
All of them are the most widely used benchmark datasets, which can provide a fair comparison of the performance and generalization capability of each model.

\begin{table*}
   \scriptsize
   \centering
   \setlength{\tabcolsep}{3.8mm}
   \renewcommand\arraystretch{1}
   \caption{Quantitative comparisons with state-of-the-art SR methods.
     The best results are \textbf{highlighted} and the second best results are \underline{underlined}.
     Obviously, MDCN achieves competitive results on all benchmark datasets.}
     \begin{tabular}{lcccccc|c}
     \hline
     \multicolumn{1}{c}{Algorithm} & Scale & \begin{tabular}[c]{@{}c@{}}Set5~\cite{bevilacqua2012}\\PSNR / SSIM\end{tabular} & \begin{tabular}[c]{@{}c@{}}Set14~\cite{zeyde2010single}\\PSNR / SSIM\end{tabular} & \begin{tabular}[c]{@{}c@{}}BSDS100~\cite{arbelaez2011}\\PSNR / SSIM\end{tabular} & \begin{tabular}[c]{@{}c@{}}Urban100~\cite{huang2015singl}\\PSNR / SSIM\end{tabular} & \begin{tabular}[c]{@{}c@{}}Manga109~\cite{matsui2017sketch}\\PSNR / SSIM\end{tabular} & \begin{tabular}[c]{@{}c@{}}Average\\PSNR / SSIM\end{tabular} \\ \hline \hline

     Bicubic & $\times 2$ & 33.66 / 0.9299 & 30.24 / 0.8688 & 29.56 / 0.8431 & 26.88 / 0.8403 & 30.80 / 0.9339 & 30.22 / 0.8832\\
     A+~\cite{timofte2014a+} (2014)      & $\times 2$ & 36.60 / 0.9542 & 32.42 / 0.9059 & 31.24 / 0.8870 & 29.25 / 0.8955 & 35.37 / 0.9663 & 32.98 / 0.9218\\
     SelfExSR~\cite{huang2015single} (2015) & $\times 2$ & 36.60 / 0.9537 & 32.46 / 0.9051 & 31.20 / 0.8863 & 29.55 / 0.8983 & 35.82 / 0.9671 & 33.13 / 0.9221\\
     SRCNN~\cite{dong2014learning} (2014)   & $\times 2$ & 36.66 / 0.9542 & 32.45 / 0.9067 & 31.36 / 0.8879 & 29.50 / 0.8946 & 35.60 / 0.9663 & 33.11 / 0.9219 \\
     ESPCN~\cite{shi2016real} (2016) & $\times 2$ & 37.00 / 0.9559 & 32.75 / 0.9098 & 31.51 / 0.8939 & 29.87 / 0.9065 & 36.21 / 0.9694 & 33.47 / 0.9271  \\
     FSRCNN~\cite{dong2016accelerating} (2016) & $\times 2$ & 37.06 / 0.9554 & 32.76 / 0.9078 & 31.53 / 0.8912 & 29.88 / 0.9024 & 36.67 / 0.9694 &  33.58 / 0.9252 \\
     VDSR~\cite{kim2016accurate} (2016)  & $\times 2$ & 37.53 / 0.9590 & 33.05 / 0.9130 &  31.90 / 0.8960 &  30.77 / 0.9140 & 37.22 / 0.9750 & 34.09 / 0.9314\\
     DRCN~\cite{kim2016deeply}(2016)  & $\times 2$ &  37.63 / 0.9584 & 33.06 / 0.9108 &  31.85 / 0.8947 &  30.76 / 0.9147 &  37.63 / 0.9723 & 34.19 / 0.9302 \\
     LapSRN~\cite{lai2017deep} (2017) & $\times 2$ & 37.52 / 0.9591 &  33.08 / 0.9130 & 31.80 / 0.8950 & 30.41 / 0.9101 &  37.27 / 0.9740 & 34.02 / 0.9302\\
     EDSR~\cite{lim2017enhanced} (2017) & $\times 2$  & 38.11 / 0.9602 & 33.92 / 0.9195 & 32.32 / 0.9013 & \underline{32.93} / 0.9351 & 39.10 / 0.9773 & 35.27 / 0.9387\\
     SRMDNF~\cite{zhang2018learning} (2018) & $\times 2$ & 37.79 / 0.9601 & 33.32 / 0.9159 & 32.05 / 0.8985 & 31.33 / 0.9204 & 38.07 / 0.9761 & 34.51 / 0.9342 \\
     MSRN~\cite{Li_2018_ECCV} (2018)   & $\times 2$ & 38.07 / 0.9608 & 33.68 / 0.9184 & 32.22 / 0.9002 & 32.32 / 0.9304 & 38.64 / 0.9771  & 34.99 / 0.9374 \\
     RDN~\cite{zhang2018residual} (2018)  & $\times 2$  &  \underline{38.24 / 0.9614} & \underline{34.01 / 0.9212} & \underline{32.34 / 0.9017} & 32.89 / 0.9353 & \underline{39.18 / 0.9780} & \underline{35.33 / 0.9395} \\
     RAN~\cite{Wang2019ResolutionAwareNF} (2019) & $\times 2$ & 37.58 / 0.9592 & 33.10 / 0.9133 & 31.92 / 0.8963 & N / A & N / A & N / A \\
     DNCL~\cite{Xie2019FastSS} (2019) & $\times 2$  & 37.65 / 0.9599 & 33.18 / 0.9141 & 31.97 / 0.8971 & 30.89 / 0.9158 & N / A & N / A \\
     FilterNet~\cite{Li2019FilterNetAI} (2019) & $\times 2$ & 37.86 / 0.9610 &  33.34 / 0.9150 & 32.09 / 0.8990 & 31.24 / 0.9200 &  N / A & N / A \\
     MRFN~\cite{he2019mrfn} (2019) & $\times 2$ & 37.98 / 0.9611 & 33.41 / 0.9159 & 32.14 / 0.8997 & 31.45 / 0.9221 & 38.29 / 0.9759 & 34.65 / 0.9349 \\
     SeaNet~\cite{fang2020soft} (2020) & $\times 2$  &  38.08 / 0.9609 & 33.75 / 0.9190 & 32.27 / 0.9008 & 32.50 / 0.9318 & 38.76 / 0.9774  & 35.07 / 0.9380 \\
     MDCN (Ours) & $\times 2$ &   38.19 / 0.9612  & 33.86 / 0.9202  & 32.32 / 0.9014  & 32.92 / \underline{0.9355}  & 39.09 / \underline{0.9780} & 35.28 / 0.9393 \\
     MDCN+ (Ours) & $\times 2$ & \textbf{38.25 / 0.9614} &  \textbf{34.01 / 0.9208}  & \textbf{32.36 / 0.9019}  &   \textbf{33.08 / 0.9367}  & \textbf{39.27 / 0.9784} & \textbf{35.39 / 0.9399} \\
     \hline
     \hline
     Bicubic & $\times 3$ & 30.39 / 0.8682	& 27.55 / 0.7742	& 27.21 / 0.7385	& 24.46 / 0.7449	& 26.95 / 0.8556	& 27.31 / 0.7963 \\
     A+~\cite{timofte2014a+} (2014)  & $\times 3$ & 32.63 / 0.9085 & 29.25 / 0.8194 & 28.31 / 0.7828 & 26.05 / 0.8019 & 29.93 / 0.9089 & 29.23 / 0.8443 \\
     SelfExSR~\cite{huang2015single} (2015) & $\times 3$ & 32.66 / 0.9089 & 29.34 / 0.8222 & 28.30 / 0.7839 & 26.45 / 0.8124 & 27.57 / 0.7997 & 28.86 / 0.8254\\
     SRCNN~\cite{dong2014learning} (2014)  & $\times 3$ & 32.75 / 0.9090 & 29.30 / 0.8215 & 28.41 / 0.7863 & 26.24 / 0.7989 & 30.48 / 0.9117 & 29.44 / 0.8455 \\
     ESPCN~\cite{shi2016real} (2016) & $\times 3$ & 33.02 / 0.9135 & 29.49 / 0.8271 & 28.50 / 0.7937 & 26.41 / 0.8161 & 30.79 / 0.9181 & 29.64 / 0.8537 \\
     FSRCNN~\cite{dong2016accelerating} (2016) & $\times 3$ & 33.20 / 0.9149 & 29.54 / 0.8277 & 28.55 / 0.7945 & 26.48 / 0.8175 & 30.98 / 0.9212 &  29.75 / 0.8552 \\
     VDSR~\cite{kim2016accurate} (2016)  & $\times 3$ & 33.67 / 0.9210 & 29.78 / 0.8320 & 28.83 / 0.7990 & 27.14 / 0.8290 & 32.01 / 0.9340 & 30.29 / 0.8630 \\
     DRCN~\cite{kim2016deeply} (2016) & $\times 3$ & 33.85 / 0.9215 & 29.89 / 0.8317 & 28.81 / 0.7954  & 27.16 / 0.8311  & 32.31 / 0.9328 & 30.40 / 0.8625 \\
     LapSRN~\cite{lai2017deep} (2017) & $\times 3$ & 33.82 / 0.9227 & 29.87 / 0.8320 & 28.82 / 0.7980  & 27.07 / 0.8280  & 32.21 / 0.9350 & 30.36 / 0.8631 \\
     EDSR~\cite{lim2017enhanced} (2017) & $\times 3$  & 34.65 / 0.9280 & 30.52 / 0.8462 & 29.25 / 0.8093 & 28.80 / 0.8653 & \underline{34.17} / 0.9476 & 31.48 / 0.8793 \\
     SRMDNF~\cite{zhang2018learning} (2018) & $\times 3$ & 34.12 / 0.9254 & 30.04 / 0.8382 & 28.97 / 0.8025 & 27.57 / 0.8398 & 33.00 / 0.9403 & 30.74 / 0.8692 \\
     MSRN~\cite{Li_2018_ECCV}  (2018) & $\times 3$ & 34.48 / 0.9276 & 30.40 / 0.8436 & 29.13 / 0.8061 & 28.31 / 0.8560 & 33.56 / 0.9451 & 31.18 / 0.8757 \\
     RDN~\cite{zhang2018residual} (2018) & $\times 3$ &  \underline{34.71 / 0.9296} & \underline{30.57 / 0.8468}  & \underline{29.26} / 0.8093 & 28.80 / 0.8653  & 34.13 / 0.9484  & 31.49 / 0.8799 \\
     RAN~\cite{Wang2019ResolutionAwareNF} (2019) & $\times 3$ & 33.71 / 0.9223 &  29.84 / 0.8326 & 28.84 / 0.7981 & N / A &  N / A & N / A \\
     DNCL~\cite{Xie2019FastSS} (2019)&  $\times 3$  & 33.95 / 0.9232 & 29.93 / 0.8340 & 28.91 / 0.7995 & 27.27 / 0.8326 & N / A  & N / A \\
     FilterNet~\cite{Li2019FilterNetAI} (2019) & $\times 3$ &  34.08 / 0.9250 & 30.03 / 0.8370 & 28.95 / 0.8030 & 27.55 / 0.8380 & N / A & N / A \\
     MRFN~\cite{he2019mrfn} (2019) & $\times 3$ & 34.21 / 0.9267 & 30.03 / 0.8363 & 28.99 / 0.8029 & 27.53 / 0.8389 & 32.82 / 0.9396 & 30.72 / 0.8689 \\
     SeaNet~\cite{fang2020soft} (2020) & $\times 3$  &  34.55 / 0.9282 & 30.42 / 0.8444 & 29.17 / 0.8071 & 28.50 / 0.8594 & 33.73 / 0.9463 &  31.27 /  0.8771 \\
     MDCN (Ours) & $\times 3$ &  34.69 / 0.9294 &   30.54 / 0.8470  &  \underline{29.26 / 0.8095}  &  \underline{28.83 / 0.8662}  & \underline{34.17 / 0.9485} & \underline{31.50 / 0.8801} \\
     MDCN+ (Ours) & $\times 3$ &   \textbf{34.76 / 0.9299} &  \textbf{30.63 / 0.8480}  &  \textbf{29.31 / 0.8103}  &  \textbf{29.00 / 0.8687}  &  \textbf{34.43 / 0.9497} &  \textbf{31.63 / 0.8813} \\
     \hline
     \hline
     Bicubic & $\times 4$ & 28.42 / 0.8104	& 26.00 / 0.7027	& 25.96 / 0.6675	& 23.14 / 0.6577	& 24.89 / 0.7866	& 25.68 / 0.7250 \\
     A+~\cite{timofte2014a+} (2014)     & $\times 4$ & 30.33 / 0.8565 & 27.44 / 0.7450 & 26.83 / 0.6999 & 24.34 / 0.7211 & 27.03 / 0.8439 & 27.19 / 0.7733 \\
     SelfExSR~\cite{huang2015single} (2015) & $\times 4$ & 30.34 / 0.8593 & 27.55 / 0.7511 & 26.84 / 0.7032 & 24.83 / 0.7403 & 27.83 / 0.8598 & 27.48 / 0.7827 \\
     SRCNN~\cite{dong2014learning} (2014)  & $\times 4$ & 30.48 / 0.8628 & 27.50 / 0.7513 & 26.90 / 0.7101 & 24.52 / 0.7221 & 27.58 / 0.8555 & 27.40 / 0.7804 \\
     ESPCN~\cite{shi2016real} (2016) & $\times 4$ & 30.66 / 0.8646 & 27.71 / 0.7562 & 26.98 / 0.7124 & 24.60 / 0.7360 & 27.70 / 0.8560 & 27.53 / 0.7850 \\
     FSRCNN~\cite{dong2016accelerating} (2016) & $\times 4$ & 30.73 / 0.8601 & 27.71 / 0.7488 & 26.98 / 0.7029 & 24.62 / 0.7272 & 27.90 / 0.8517  & 27.59 / 0.7781 \\
     VDSR~\cite{kim2016accurate} (2016)   & $\times 4$ & 31.35 / 0.8830 & 28.02 / 0.7680 &  27.29 / 0.7267 &  25.18 / 0.7540 & 28.83 / 0.8870 & 28.13 / 0.8037 \\
     DRCN~\cite{kim2016deeply} (2016)   & $\times 4$ &  31.56 / 0.8810 & 28.15 / 0.7627 & 27.24 / 0.7150 &  25.15 / 0.7530 & 28.98 / 0.8816 & 28.22 / 0.7987 \\
     LapSRN~\cite{lai2017deep} (2017) & $\times 4$ &  31.54 / 0.8850 & 28.19 / 0.7720 & 27.32 / 0.7270 &  25.21 / 0.7560 & 29.09 / 0.8900 & 28.27 / 0.8060 \\
     EDSR~\cite{lim2017enhanced} (2017) & $\times 4$  & 32.46 / 0.8968 & 28.80 / 0.7876 & 27.71 / 0.7420 & 26.64 / 0.8033 & 31.02 / 0.9148 & 29.33 / 0.8289 \\
     SRMDNF~\cite{zhang2018learning} (2018) & $\times 4$  & 31.96 / 0.8925 & 28.35 / 0.7787 & 27.49 / 0.7337 & 25.68 / 0.7731 & 30.09 / 0.9024 & 28.71 / 0.8161 \\
     MSRN~\cite{Li_2018_ECCV} (2018)  & $\times 4$ & 32.25 / 0.8958 & 28.63 / 0.7833 & 27.61 / 0.7377 & 26.22 / 0.7905 & 30.57 / 0.9103 & 29.05 / 0.8235 \\
     RDN~\cite{zhang2018residual} (2018) & $\times 4$ &  32.47 /  \underline{0.8990} & 28.81 / 0.7871 & 27.72 / 0.7419 & 26.61 / 0.8028  & 31.00 / 0.9151  & 29.32 / 0.8292 \\
     RAN~\cite{Wang2019ResolutionAwareNF} (2019) & $\times 4$ & 31.43 / 0.8847 & 28.09 / 0.7691 & 27.31 / 0.7260 & N / A & N / A & N / A \\
     DNCL~\cite{Xie2019FastSS} (2019) & $\times 4$ & 31.66 / 0.8871 &  28.23 / 0.7717 & 27.39 / 0.7282 & 25.36 / 0.7606 &  N / A & N / A \\
     FilterNet~\cite{Li2019FilterNetAI}  (2019) & $\times 4$  & 31.74 / 0.8900 & 28.27 / 0.7730 & 27.39 / 0.7290 & 25.53 / 0.7680 & N / A & N / A \\
     MRFN~\cite{he2019mrfn} (2019) & $\times 4$ & 31.90 / 0.8916 & 28.31 / 0.7746 & 27.43 / 0.7309 & 25.46 / 0.7654 & 29.57 / 0.8962 & 28.53 / 0.8117 \\
     SeaNet~\cite{fang2020soft} (2020)  & $\times 4$  &  32.33 / 0.8970 & 28.72 / 0.7855 & 27.65 / 0.7388 & 26.32 / 0.7942 & 30.74 / 0.9129 &  29.13 / 0.8257 \\
     MDCN (Ours) & $\times 4$ &  \underline{32.48} / 0.8985 & \underline{28.83 / 0.7879}  & \underline{27.74 / 0.7423}  & \underline{26.69 / 0.8049}  & \underline{31.10 / 0.9163} & \underline{29.37 / 0.8300} \\
     MDCN+ (Ours) & $\times 4$ &   \textbf{32.61 / 0.9000} &  \textbf{28.90 / 0.7893}  &  \textbf{27.79 / 0.7434}  &  \textbf{26.86 / 0.8083}  &  \textbf{31.40 / 0.9188} & \textbf{29.51 / 0.8320} \\
     \hline \\
     \end{tabular}
     \label{Results}
\end{table*}

\begin{table*}
   \scriptsize
   \centering
   \setlength{\tabcolsep}{0.8mm}
   \renewcommand\arraystretch{1.1}
   \caption{Quantitative comparisons (PSNR/SSIM, Parameters, and Execution time) with RCAN~\cite{Zhang_2018_ECCV}. The best results are \textbf{highlighted} and the fastest execution time are RED.}
     \label{Results-RCAN}
   \begin{tabular}{|c|c|c|c|c|c|c|c||c|}
     \hline
     Algorithm & Scale & Parameters & \begin{tabular}[c]{@{}c@{}}Set5~\cite{bevilacqua2012}\\PSNR / SSIM / Time\end{tabular} & \begin{tabular}[c]{@{}c@{}}Set14~\cite{zeyde2010single}\\PSNR / SSIM / Time\end{tabular} & \begin{tabular}[c]{@{}c@{}}BSDS100~\cite{arbelaez2011}\\PSNR / SSIM / Time\end{tabular} & \begin{tabular}[c]{@{}c@{}}Urban100~\cite{huang2015singl}\\PSNR / SSIM / Time\end{tabular} & \begin{tabular}[c]{@{}c@{}}Manga109~\cite{matsui2017sketch}\\PSNR / SSIM / Time\end{tabular} & \begin{tabular}[c]{@{}c@{}}Average\\PSNR / SSIM / Time\end{tabular}\\ \hline \hline

     RCAN~\cite{Zhang_2018_ECCV}  & $\times 2$  & 15M &  \textbf{38.27 / 0.9614} / 0.60s & \textbf{34.12 / 0.9216} / 1.11s & \textbf{32.41 / 0.9027} / 0.75s & \textbf{33.34 / 0.9384} / 3.78s & \textbf{39.44 / 0.9786} / 4.55s & \textbf{35.52 / 0.9405} / 2.16s \\
     MDCN (Ours) & $\times 2$ &  15M &  38.19 / 0.9612 / {\color{red}0.22s}  & 33.86 / 0.9202 / {\color{red}0.38s} &  32.32 / 0.9014 / {\color{red}0.29s} & 32.92 / 0.9355 / {\color{red}1.19s} &  39.09 / 0.9780 / {\color{red}1.45s} &  35.28 / 0.9393 / {\color{red}0.71s} \\
     \hline
     \hline

     RCAN~\cite{Zhang_2018_ECCV}  & $\times 3$ & 15M &  \textbf{34.74 / 0.9299} / 0.34s & \textbf{30.65 / 0.8482} / 0.55s & \textbf{29.32 / 0.8111} / 0.41s & \textbf{29.09 / 0.8702} / 1.89s & \textbf{34.44 / 0.9499} / 2.33s & \textbf{31.65 / 0.8818} / 1.10s \\
     MDCN (Ours) & $\times 3$ & 15M & 34.69 / 0.9294 / {\color{red}0.15s} &  30.54 / 0.8470 / {\color{red}0.24s} &  29.26 / 0.8095 / {\color{red}0.16s} &  28.83 / 0.8662 / {\color{red}0.71s} & 34.17 / 0.9485 / {\color{red}0.87s} & 31.50 / 0.8801 / {\color{red}0.43s}\\
     \hline
     \hline
     
     RCAN~\cite{Zhang_2018_ECCV}  & $\times 4$ & 15M &  \textbf{32.63 / 0.9002} / 0.30s & \textbf{28.87 / 0.7889} / 0.40s & \textbf{27.77 / 0.7436} / 0.30s & \textbf{26.82 / 0.8087} / 1.21s & \textbf{31.22 / 0.9173} / 1.50s & \textbf{29.46 / 0.8317} / 0.74s\\
     MDCN (Ours) & $\times 4$  & 15M &  32.48 / 0.8985 / {\color{red}0.12s} & 28.83 / 0.7879 / {\color{red}0.18s}  & 27.74 / 0.7423 / {\color{red}0.12s} & 26.69 / 0.8049 / {\color{red}0.52s} & 31.10 / 0.9163 / {\color{red}0.62s} & 29.37 / 0.8300 / {\color{red}0.31s}\\
     \hline
     \end{tabular}
     \setlength{\belowcaptionskip}{-0.9cm}
 \end{table*}
\subsection{Implementation Details}
\textbf{Model setting:}
In the final version of MDCN, we use 12 MDCBs ($N = 12$) for feature extraction, the input and output channel of each MDCB are set as 128 ($C = 128$).
Detailed configuration of the input and output channels for each layer in each module has been provided in Fig.~\ref{MDCB} and Fig.~\ref{Distillation}.
We also introduce the self-ensemble mechanism~\cite{timofte2016seven} to further improve MDCN, which is denoted as \textbf{MDCN+}.
Specifically, we first flip and rotate the input image to generate seven augmented inputs for each sample, thereby, we can obtain eight corresponding inputs.
Then, we reconstruct the corresponding SR images using our MDCN.
Finally, we apply the inverse transformation to those SR images and average them to generate the final SR image.

\begin{figure*}[h]
   \begin{minipage}[c]{0.13\textwidth}
   \includegraphics[width=2.3cm]{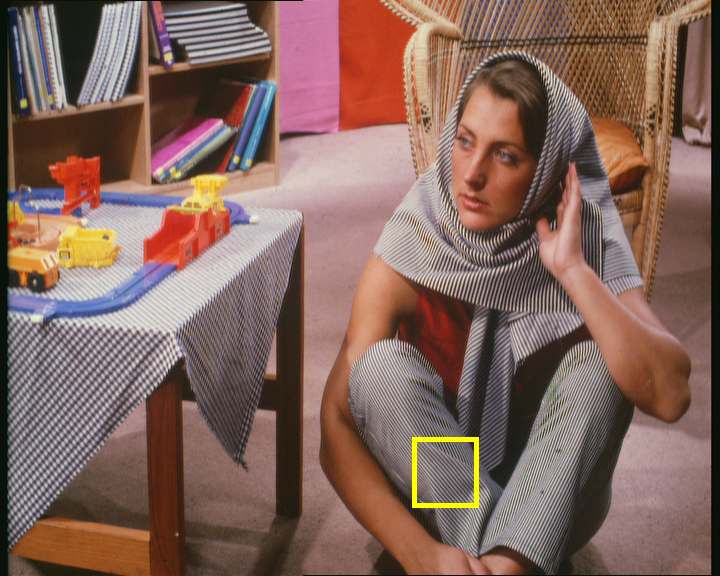}
   \centerline{$\times$2: barbara~\cite{zeyde2010single}}
   \end{minipage}
   \begin{minipage}[c]{0.11\textwidth}
   \includegraphics[width=2cm, height=1.75cm]{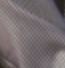}
   \centerline{Bicubic}
   \includegraphics[width=2cm, height=1.75cm]{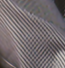}
   \centerline{MSRN}
   \end{minipage}
   \begin{minipage}[c]{0.11\textwidth}
   \includegraphics[width=2cm, height=1.75cm]{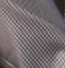}
   \centerline{A+}
   \includegraphics[width=2cm, height=1.75cm]{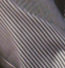}
   \centerline{EDSR}
   \end{minipage}
   \begin{minipage}[c]{0.11\textwidth}
   \includegraphics[width=2cm, height=1.75cm]{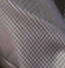}
   \centerline{SelfExSR}
   \includegraphics[width=2cm, height=1.75cm]{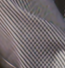}
   \centerline{SeaNet}
   \end{minipage}
   \begin{minipage}[c]{0.11\textwidth}
   \includegraphics[width=2cm, height=1.75cm]{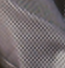}
   \centerline{SRCNN}
   \includegraphics[width=2cm, height=1.75cm]{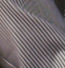}
   \centerline{RDN}
   \end{minipage}
   \begin{minipage}[c]{0.11\textwidth}
   \includegraphics[width=2cm, height=1.75cm]{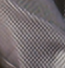}
   \centerline{FSRCNN}
   \includegraphics[width=2cm, height=1.75cm]{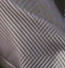}
   \centerline{RCAN}
   \end{minipage}
   \begin{minipage}[c]{0.11\textwidth}
   \includegraphics[width=2cm, height=1.75cm]{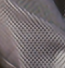}
   \centerline{LAPSRN}
   \includegraphics[width=2cm, height=1.75cm]{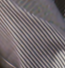}
   \centerline{MDCN (Ours)}
   \end{minipage}
   \begin{minipage}[c]{0.11\textwidth}
   \includegraphics[width=2cm, height=1.75cm]{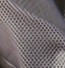}
   \centerline{DRCN}
   \includegraphics[width=2cm, height=1.75cm]{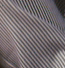}
   \centerline{Ground Truth}
   \end{minipage}

   \begin{minipage}[c]{0.13\textwidth}
   \includegraphics[width=2.3cm]{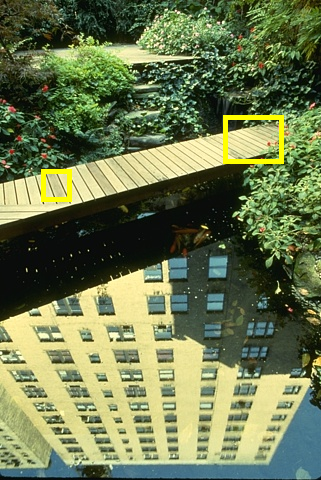}
   \centerline{$\times$3: 148026~\cite{arbelaez2011}}
   \end{minipage}
   \begin{minipage}[c]{0.11\textwidth}
   \includegraphics[width=2cm, height=1.75cm]{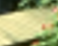}
   \centerline{Bicubic}
   \includegraphics[width=2cm, height=1.75cm]{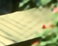}
   \centerline{MSRN}
   \end{minipage}
   \begin{minipage}[c]{0.11\textwidth}
   \includegraphics[width=2cm, height=1.75cm]{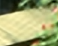}
   \centerline{A+}
   \includegraphics[width=2cm, height=1.75cm]{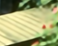}
   \centerline{EDSR}
   \end{minipage}
   \begin{minipage}[c]{0.11\textwidth}
   \includegraphics[width=2cm, height=1.75cm]{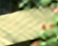}
   \centerline{SelfExSR}
   \includegraphics[width=2cm, height=1.75cm]{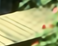}
   \centerline{SeaNet}
   \end{minipage}
   \begin{minipage}[c]{0.11\textwidth}
   \includegraphics[width=2cm, height=1.75cm]{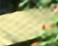}
   \centerline{SRCNN}
   \includegraphics[width=2cm, height=1.75cm]{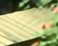}
   \centerline{RDN}
   \end{minipage}
   \begin{minipage}[c]{0.11\textwidth}
   \includegraphics[width=2cm, height=1.75cm]{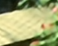}
   \centerline{FSRCNN}
   \includegraphics[width=2cm, height=1.75cm]{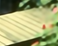}
   \centerline{RCAN}
   \end{minipage}
   \begin{minipage}[c]{0.11\textwidth}
   \includegraphics[width=2cm, height=1.75cm]{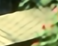}
   \centerline{LAPSRN}
   \includegraphics[width=2cm, height=1.75cm]{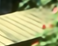}
   \centerline{MDCN (Ours)}
   \end{minipage}
   \begin{minipage}[c]{0.11\textwidth}
   \includegraphics[width=2cm, height=1.75cm]{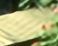}
   \centerline{DRCN}
   \includegraphics[width=2cm, height=1.75cm]{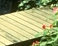}
   \centerline{Ground Truth}
   \end{minipage}

   \begin{minipage}[c]{1\textwidth}
   \end{minipage}
\caption{Visual comparison with different SR methods on Set14~\cite{zeyde2010single} or BSDS100~\cite{arbelaez2011} under small upsampling factors. \textbf{Please zoom in to view details.}}
\label{Visual-1}
\end{figure*}

\begin{figure*}[h]
   \begin{minipage}[c]{0.13\textwidth}
   \includegraphics[width=2.3cm]{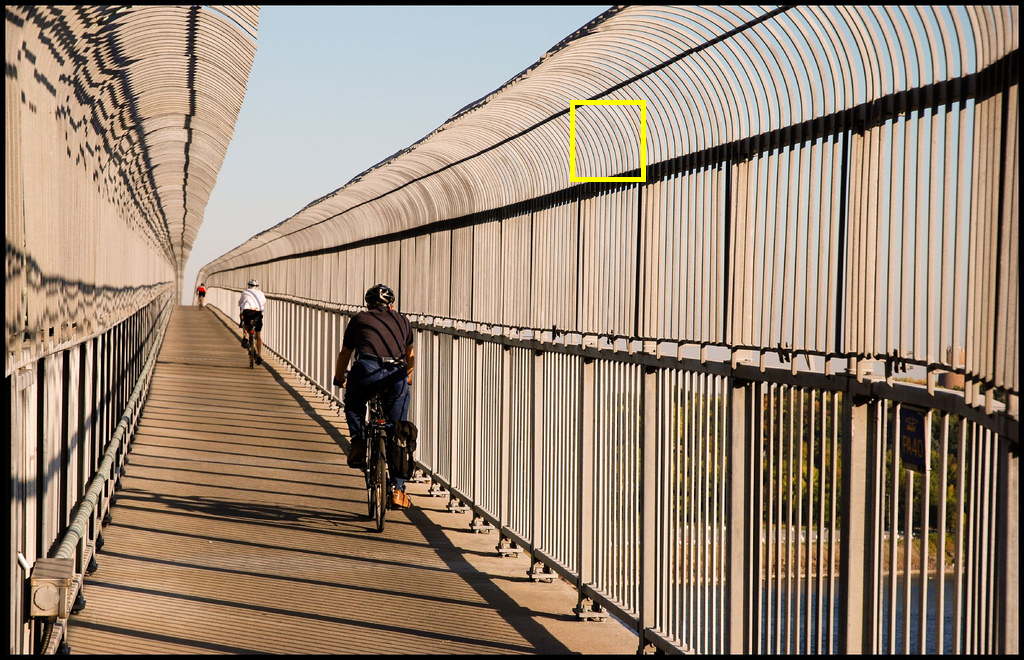}
   \centerline{$\times$4: img\_024~\cite{huang2015singl}}
   \end{minipage}
   \begin{minipage}[c]{0.11\textwidth}
   \includegraphics[width=2cm, height=1.75cm]{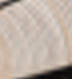}
   \centerline{Bicubic}
   \includegraphics[width=2cm, height=1.75cm]{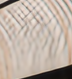}
   \centerline{MSRN}
   \end{minipage}
   \begin{minipage}[c]{0.11\textwidth}
   \includegraphics[width=2cm, height=1.75cm]{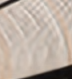}
   \centerline{A+}
   \includegraphics[width=2cm, height=1.75cm]{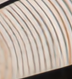}
   \centerline{EDSR}
   \end{minipage}
   \begin{minipage}[c]{0.11\textwidth}
   \includegraphics[width=2cm, height=1.75cm]{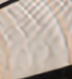}
   \centerline{SelfExSR}
   \includegraphics[width=2cm, height=1.75cm]{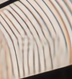}
   \centerline{SeaNet}
   \end{minipage}
   \begin{minipage}[c]{0.11\textwidth}
   \includegraphics[width=2cm, height=1.75cm]{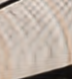}
   \centerline{SRCNN}
   \includegraphics[width=2cm, height=1.75cm]{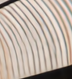}
   \centerline{RDN}
   \end{minipage}
   \begin{minipage}[c]{0.11\textwidth}
   \includegraphics[width=2cm, height=1.75cm]{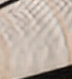}
   \centerline{FSRCNN}
   \includegraphics[width=2cm, height=1.75cm]{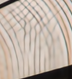}
   \centerline{RCAN}
   \end{minipage}
   \begin{minipage}[c]{0.11\textwidth}
   \includegraphics[width=2cm, height=1.75cm]{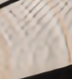}
   \centerline{LAPSRN}
   \includegraphics[width=2cm, height=1.75cm]{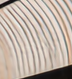}
   \centerline{MDCN (Ours)}
   \end{minipage}
   \begin{minipage}[c]{0.11\textwidth}
   \includegraphics[width=2cm, height=1.75cm]{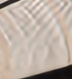}
   \centerline{DRCN}
   \includegraphics[width=2cm, height=1.75cm]{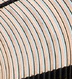}
   \centerline{Ground Truth}
   \end{minipage}

   \begin{minipage}[c]{0.13\textwidth}
   \includegraphics[width=2.3cm]{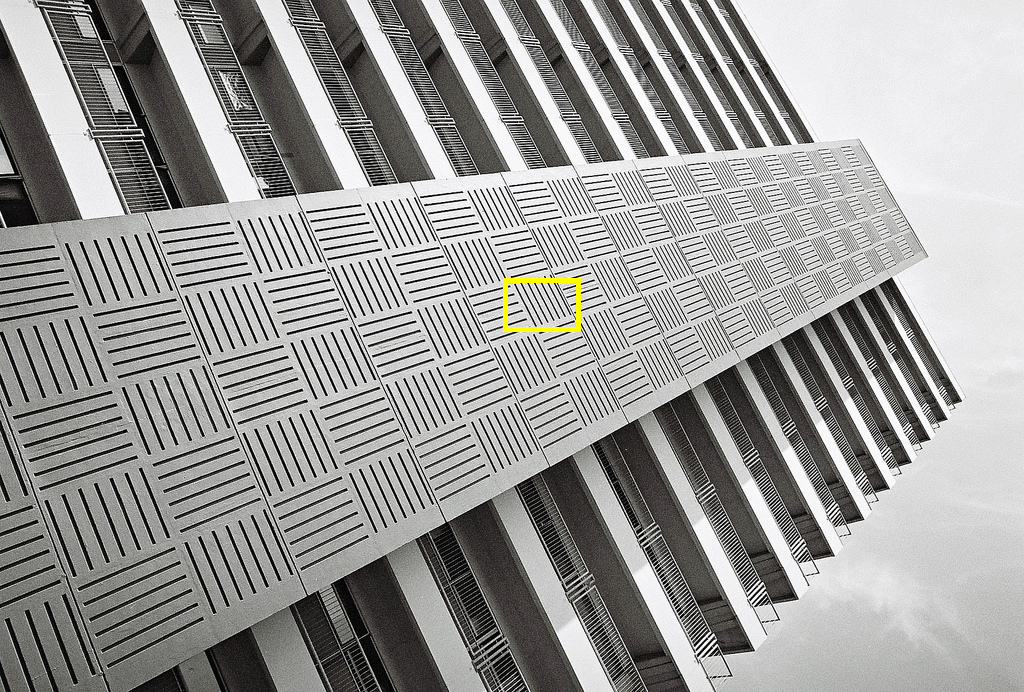}
   \centerline{$\times$4: img\_092~\cite{arbelaez2011}}
   \end{minipage}
   \begin{minipage}[c]{0.11\textwidth}
   \includegraphics[width=2cm, height=1.75cm]{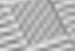}
   \centerline{Bicubic}
   \includegraphics[width=2cm, height=1.75cm]{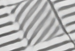}
   \centerline{MSRN}
   \end{minipage}
   \begin{minipage}[c]{0.11\textwidth}
   \includegraphics[width=2cm, height=1.75cm]{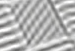}
   \centerline{A+}
   \includegraphics[width=2cm, height=1.75cm]{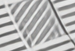}
   \centerline{EDSR}
   \end{minipage}
   \begin{minipage}[c]{0.11\textwidth}
   \includegraphics[width=2cm, height=1.75cm]{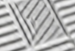}
   \centerline{SelfExSR}
   \includegraphics[width=2cm, height=1.75cm]{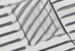}
   \centerline{SeaNet}
   \end{minipage}
   \begin{minipage}[c]{0.11\textwidth}
   \includegraphics[width=2cm, height=1.75cm]{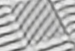}
   \centerline{SRCNN}
   \includegraphics[width=2cm, height=1.75cm]{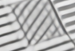}
   \centerline{RDN}
   \end{minipage}
   \begin{minipage}[c]{0.11\textwidth}
   \includegraphics[width=2cm, height=1.75cm]{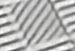}
   \centerline{FSRCNN}
   \includegraphics[width=2cm, height=1.75cm]{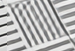}
   \centerline{RCAN}
   \end{minipage}
   \begin{minipage}[c]{0.11\textwidth}
   \includegraphics[width=2cm, height=1.75cm]{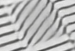}
   \centerline{LAPSRN}
   \includegraphics[width=2cm, height=1.75cm]{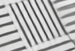}
   \centerline{MDCN (Ours)}
   \end{minipage}
   \begin{minipage}[c]{0.11\textwidth}
   \includegraphics[width=2cm, height=1.75cm]{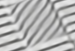}
   \centerline{DRCN}
   \includegraphics[width=2cm, height=1.75cm]{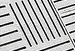}
   \centerline{Ground Truth}
   \end{minipage}

   \begin{minipage}[c]{1\textwidth}
   \end{minipage}
\caption{Visual comparison with different SR methods on Urban100~\cite{huang2015singl} under large upsampling factor ($\times 4$). \textbf{Please zoom in to view details.}}
\label{Visual-2}
\end{figure*}

\textbf{Training setting:}
During training, we use RGB image as input and augment the image with random horizontal flips and rotations.
It is worth noting that our MDCN is a multi-factor model.
Therefore, we use the mix training method to train our model.
During training, we extract 16 LR patches with a randomly selected scale among $\times$2, $\times$3, and $\times$4, 1,000 iterations of back-propagation constitute an epoch.
The learning rate is initialized to $10^{-4}$ and halved every 200 epochs.
Different from previous works~\cite{Li_2018_ECCV, Zhang_2018_ECCV} which set the size of LR images as 48$\times$48, we set the size of HR images as 48$\times$48.
Thence, the size of the corresponding LR images of different upsampling factors is 24$\times$24, 16$\times$16 and 12$\times$12, respectively.
Using small patch size for training will cause performance degradation, but greatly reduce the training time.
However, our MDCN still achieves great results due to the fully use of the inter-scale correlation between different upsampling factors.
We implement our model with the PyTorch framework and update it with Adam optimizer, all our experiments are performed on GTX TitanX.

\subsection{Comparisons with state-of-the-art SR methods}
We compare MDCN with more than 19 SR methods, including Bicubic, A+~\cite{timofte2014a+}, SelfExSR~\cite{huang2015single}, SRCNN~\cite{dong2014learning}, ESPCN~\cite{shi2016real}, FSRCNN~\cite{dong2016accelerating}, VDSR~\cite{kim2016accurate}, LapSRN~\cite{lai2017deep}, DRCN~\cite{kim2016deeply}, MRFN~\cite{he2019mrfn}, SRMDNF~\cite{zhang2018learning}, MSRN~\cite{Li_2018_ECCV}, EDSR~\cite{lim2017enhanced}, RDN~\cite{zhang2018residual}, RCAN~\cite{Zhang_2018_ECCV}, FilterNet~\cite{Li2019FilterNetAI}, DNCL~\cite{Xie2019FastSS}, RAN~\cite{Wang2019ResolutionAwareNF}, and SeaNet~\cite{fang2020soft}.
All SR images are evaluated with PSNR and SSIM~\cite{wang2004image} on the Y channel in YCbCr space.

\textbf{Quantitative Comparison:}
In TABLE~\ref{Results}, we show the quantitative comparisons with some advanced SR methods, all of them have achieved competitive results at the time.
Among them, best results are highlighted and the second best results are underlined.
Besides, the 'Average' denotes the average results of these 5 test datasets.
Obviously, our MDCN achieves competitive results on all upsampling factors.
Among them, RDN is slightly better than MDCN under small upsampling factors ($\times$2). 
However, our MDCN can achieve better results under large upsampling factors (e.g., $\times$3, $\times$4). 
This is because the introduced MDCB in MDCN can fully extract multi-scale features, which is conducive to large upsampling factor SR reconstruction.
Considering that RCAN achieves the state-of-the-out results, we make a detailed comparison with it in TABLE~\ref{Results-RCAN}. 
We can observe that RCAN is slightly better than MDCN ($\times$2: 0.24dB, $\times$3: 0.15dB, and $\times$4: 0.09dB).
However, it should be noticed that the execution time of our MDCN is 3 times faster than RCAN.
This means that MDCN can achieve similar results as RCAN with less execution time.
Meanwhile, it is worth noting that:
(1). Except for MDCN, all reported SR methods are specially trained for different upsampling factor;
(2). EDSR~\cite{lim2017enhanced}, RDN~\cite{zhang2018residual}, and RCAN~\cite{Zhang_2018_ECCV} use the pre-trained model ($\times$2) as the initialization model to train large upsampling factor model like $\times$4;
(3). DRCN~\cite{kim2016deeply} introduces the recursive mechanism to further improve model performance.
(4). Other models use large LR images as inputs for training.
All these strategies can further boost the performance.
However, in order to verify the effectiveness of MDCN, we do not use any training tricks in our experiment.
Nevertheless, since MDCB, HFDB, and DRB can extract rich image features and learn the inter-scale correlation, our MDCN still achieves competitive results.

\textbf{Visual Comparison:}
In Figs.~\ref{Visual-1} and ~\ref{Visual-2}, we show visual comparisons under small ($\times$2, $\times$3) and large upsampling factor ($\times$4), respectively.
Among them, EDSR~\cite{lim2017enhanced} was the champion model of the NTIRE2017 SR Challenge, RDN~\cite{zhang2018residual} and RCAN~\cite{Zhang_2018_ECCV} were superior models which achieved SOTA results.
According to the figure, we can clearly observe that:
(i). Most compared SR methods (e.g., SRCNN, MSRN, and SeaNet) cannot recover clear and accurate image edges.
Furthermore, under large upsample factor (e.g., $\times$4), the reconstructed SR images are blurred with severe artifacts and incorrect edges.
In contrast, our MDCN can reconstruct more realistic SR images with clear and sharp edges;
(ii). Compared with large size models (e.g., EDSR, RDN, and RCAN), our MDCN still shows competitive performance and better results in edges reconstruction.
Overall, with the help of MDCB and HFDB, MDCN can reconstruct high-quality SR images.

\section{Model Analysis}\label{Abla}
\subsection{Study of Multi-scale Dense Cross Block (MDCB)}
MDCB is the most important component of MDCN, which is designed for local and multi-scale feature extraction.

(1). As mentioned in Sec.~\ref{Method-MDCB}, MDCB is essentially a dual-path dense network, which introduces multi-scale learning, feature exchange \& fusion mechanisms, and residual learning.
In TABLE~\ref{study}, we provide a series of ablation studies to investigate their effectiveness.

\textbf{Dual-path Dense Network:}
In Cases 1 and 2, we remove all introduced mechanisms, only leaving  DenseNet-Top or DenseNet-Bottom, respectively.
In Case 3, we use both DenseNet-Top and DenseNet-Bottom to build the dual-path network for image reconstruction.
It is worth noting that Case 3 is a simple combination of these two dense networks and does not introduce skip connections ($M_{35}$ and $M_{53}$) for feature exchange and fusion.
According to the table, we can clearly see that Case 3 achieves better results, which demonstrates the dual-path network is effective.

\begin{table}
   \tiny
   \centering
   \setlength{\tabcolsep}{1.7mm}
   \renewcommand\arraystretch{1}
   \caption{Study of MDCN. 'RL' denotes 'residual learning', 'FEFM' denotes 'feature exchange \& fusion mechanism'.
   All of these models were trained under the same experimental conditions and tested on DIV2K(896-900).}
   \begin{tabular}{|c|c|c|c|c|c|c|c|c|c|}
   \hline
   \multicolumn{2}{|c|}{Case Index}                      & 1 & 2 & 3 & 4 & 5 & 6 & 7 & 8 \\ \hline \hline
   \multirow{4}{*}{MDCB}           & RL                  & $\times$  & $\times$  & $\times$   & $\times$ & $\surd$   & $\surd$  &  $\surd$  &  $\surd$   \\ \cline{2-10}
                                   & FEFM                 & $\times$  & $\times$  & $\times$   & $\surd$  & $\surd$   & $\surd$  &  $\surd$  &  $\surd$   \\ \cline{2-10}
                                   & DenseNet-Top        & $\surd$   & $\times$  & $\surd$    & $\surd$  & $\surd$   & $\surd$  &  $\surd$  &  $\surd$   \\ \cline{2-10}
                                   & DenseNet-Bottom       & $\times$  & $\surd$   & $\surd$    & $\surd$  & $\surd$   & $\surd$  &  $\surd$  &  $\surd$   \\ \hline
   \multicolumn{2}{|c|}{HFDB}                            & $\times$  & $\times$  & $\times$   & $\times$ & $\times$  & $\surd$  &  $\surd$  &  $\surd$   \\ \hline
   \multicolumn{2}{|c|}{MDCB Number}                     & 3  & 3  & 3  &  3 & 3  & 3  & 6   &  6 \\ \hline
   \multicolumn{2}{|c|}{Channel Number}                  & 128  & 128  & 128  &  128 & 128   & 128  & 128  &  256 \\ \hline
   \multicolumn{2}{|c|}{PSNR(dB) $\times$3} &  32.85 &  32.90 & 33.02  & 33.08 & 33.10 & 33.17 & \underline{33.35} & \textbf{33.49}  \\ \hline
   \end{tabular}
   \label{study}
\end{table}

\textbf{Feature Exchange \& Fusion Mechanism (FEFM):}
We introduce skip connections ($M_{35}$ and $M_{53}$) in Case 4 to achieve feature exchange and fusion.
Compared with Case 3, Case 4 achieves better results.
This is because the introduced FEFM can transfer features of different scales to the other part, which greatly enrich the diversity of extracted features.
Therefore, the model can achieve better results.

\textbf{Residual Learning (RL):}
Plenty of previous works have proved the effectiveness of residual learning.
Following these works, we also introduce residual learning into our MDCN to improve the information flow.
In Cases 4 and 5, we provide the results of the model without and with residual learning, respectively.
According to the results, we confirm the effectiveness of residual learning.
Although the improvement of PSNR is not obvious, this is because we only use 3 MDCBs here.
Furthermore, residual learning can accelerate model convergence and solve the problem of gradient disappearance and explosion, which is beneficial for deep networks.

\textbf{MDCB and Channel Number:}
MDCN is a modular network, so the size of the model can be easily adjusted by changing the number of MDCB ($N$) and channels ($C$).
In Cases 6, 7, and 8, we provide the results of different numbers of MDCBs and channels.
Obviously, as the number of MDCBs and channel increases, the model performance can be further improved.
This means the reported results are not the best, and its performance still has room for improvement.
However, in order to achieve a well balance between performance and model size, we set $N=12$ and $C=128$ in the final model.

\begin{figure}
   \centering
   \begin{minipage}[c]{1\textwidth}
   \includegraphics[width=8.8cm, height=4cm]{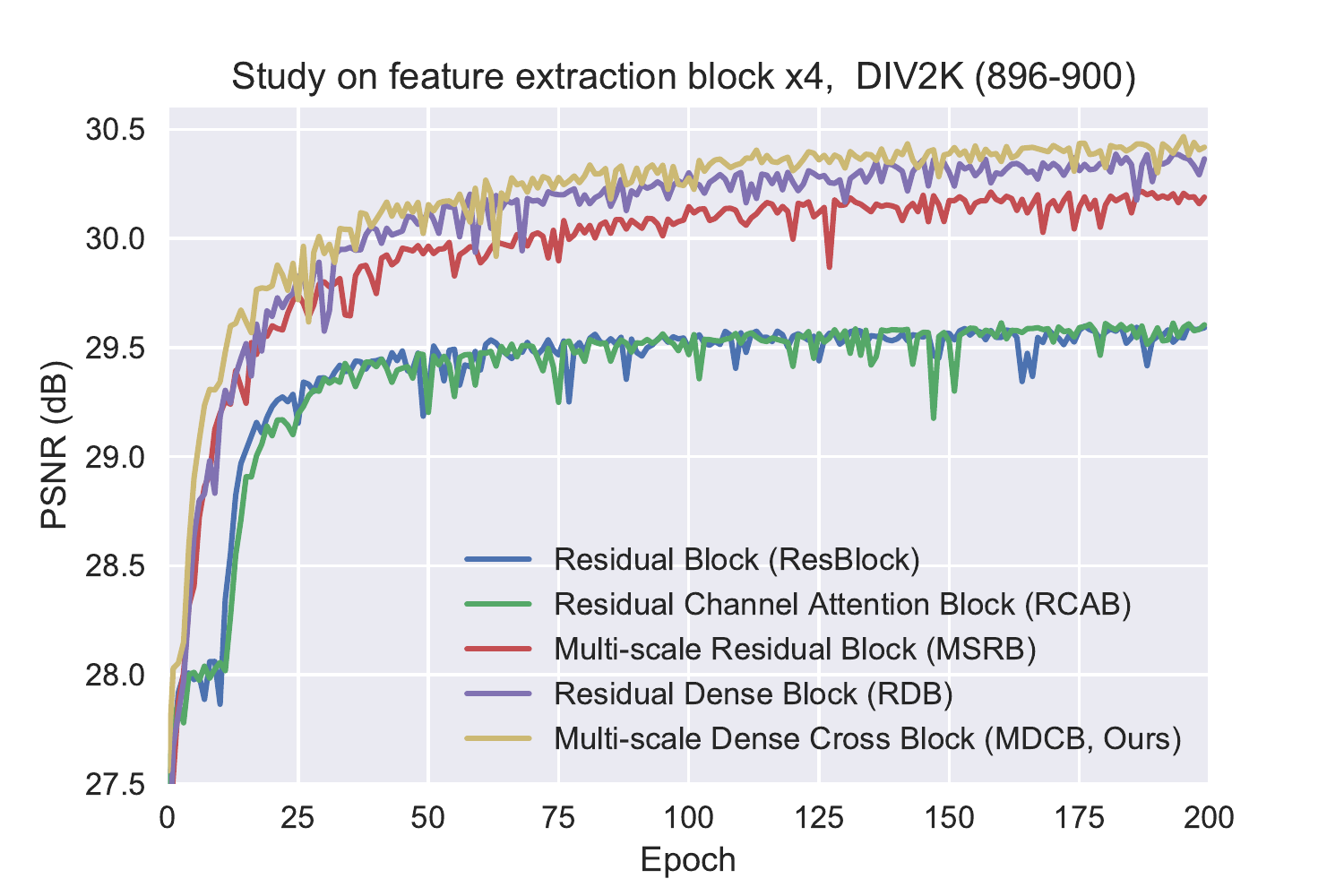}
   \end{minipage}
   \caption{Comparison of different feature extraction blocks.}
   \label{curve}
\end{figure}

(2).To further demonstrate the effectiveness of MDCB, we compare our MDCB with ResBlock\cite{he2016deep}, RDB~\cite{zhang2018residual}, RCAB~\cite{Zhang_2018_ECCV}, and MSRB~\cite{Li_2018_ECCV}.
All of them are the most widely used feature extraction blocks, which have well-designed architecture and have been widely validated.
In the experiment, we use MDCN as the basic backbone and replace the original MDCB with different feature extraction blocks (e.g., RDB and RCAB) to build new models. 
After that, all these models are retrained and tested under the same equipment, environment, and dataset. 
Meanwhile, all of these feature extraction blocks are adjusted to the same parameter level by adjusting the number of blocks or channels for a fair comparison. 
In Fig.~\ref{curve}, we show the performance curve of these models during training.
According to this figure, we can draw the following conclusions:
(i) simply stacking ResBlock or RCAB will not effectively boost model performance;
(ii) compared with MSRB, the result of MDCB has been significantly improved;
(iii) the performance of MDCB is comparable to RDB, even slightly better than it.
This is because MDCB skillfully combines multi-scale residual learning with dense connections to make it have the characteristics of both MSRB and RDB, so as to obtain better results.
All aforementioned experiments fully demonstrate the effectiveness of MDCB. 

\subsection{Study of Hierarchical Feature Distillation Block (HFDB)}
\begin{figure}
   \centering
   \begin{minipage}[c]{1\textwidth}
   \includegraphics[width=8.8cm, height=4cm]{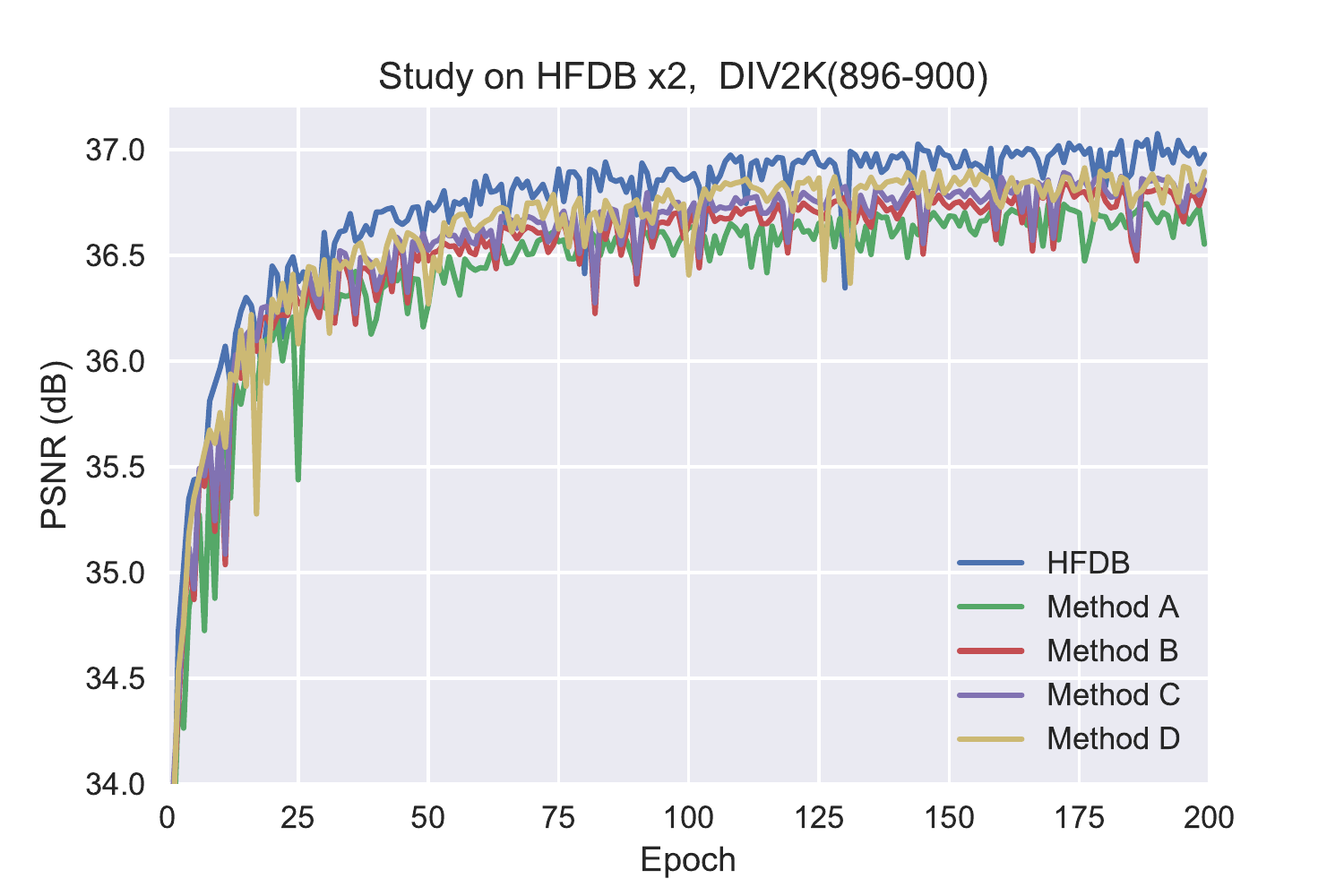}
   \end{minipage}
   \caption{Comparison of different hierarchical feature utilization methods.}
   \label{HFDB-curve}
\end{figure}

In order to eliminate redundant features and extract the most useful hierarchical features, we propose HFDB.
In TABLE~\ref{study}, Cases 5 and 6 represent the results of the model without and with HFDB, respectively.
Obviously, with the help of HFDB, the model performance can be further improved.
On the other hand, we compare HFDB with four hierarchical feature utilization methods mentioned in Fig.~\ref{Connection}.
To ensure the fairness of comparison, we apply these methods to the same model and the performance curves of each method are presented in Fig.~\ref{HFDB-curve}. 
Obviously, our HFDB achieves the best results (the blue line).
It is worth noting that the introduced dimension transformation mechanism in HFDB can greatly reduce model parameters.
Taking 12 MDCBs as an example, Method D needs $245,760$ parameters while our HFDB only needs $173,184$ parameters, $5/7$ of Method D.
In addition, this gap will increase as the number of MDCB increases.
In summary, compared with Method D (proposed in MSRN~\cite{Li_2018_ECCV}), our HFDB can achieve better results with fewer parameters and less computational overhead.

\begin{table}
   \tiny
   \centering
   \setlength{\tabcolsep}{1.3mm}
   \renewcommand\arraystretch{1.2}
   \caption{Quantitative comparisons with MSRN, MDCN-S, and MDCN.}
   \begin{tabular}{lcccccc}
   \hline
   \multicolumn{1}{c}{Algorithm} & Scale & \begin{tabular}[c]{@{}c@{}}Set5~\cite{bevilacqua2012}\\PSNR / SSIM\end{tabular} & \begin{tabular}[c]{@{}c@{}}Set14~\cite{zeyde2010single}\\PSNR / SSIM\end{tabular} & \begin{tabular}[c]{@{}c@{}}BSDS100~\cite{arbelaez2011}\\PSNR / SSIM\end{tabular} & \begin{tabular}[c]{@{}c@{}}Urban100~\cite{huang2015singl}\\PSNR / SSIM\end{tabular} & \begin{tabular}[c]{@{}c@{}}Manga109~\cite{matsui2017sketch}\\PSNR / SSIM\end{tabular} \\ \hline
   MSRN~\cite{Li_2018_ECCV}   & $\times 4$ & 32.25 / 0.8958 & 28.63 / 0.7833 & 27.61 / 0.7377 & 26.22 / 0.7905 & 30.57 / 0.9103 \\
   MDCN-S                    & $\times 4$ & 32.39 / 0.8977 & 28.75 / 0.7863 & 27.66 / 0.7398 & 26.45 / 0.7985 & 30.87 / 0.9141 \\
   MDCN                & $\times 4$ &  \textbf{32.48 / 0.8985} & \textbf{28.83 / 0.7879}  & \textbf{27.74 / 0.7423}  & \textbf{26.69 / 0.8049}  & \textbf{31.10 / 0.9163}  \\ \hline
   \end{tabular}
   \label{MDCN_S}
\end{table}

\subsection{Study of Dynamic Reconstruction Block (DRB)} \label{SDRB}
To realize the reconstruction of SR images with different upsampling factors in a single model, we introduce DRB into MDCN.
DRB can create a set of scale-specific upsampling modules according to actual needs.
This maximizes the reuse of model parameters and enables model to learn the inter-scale correlation between different upsampling factors.
It is worth noting that when there is only one upsampling module in DRB, it is same as previous works that specifically trained for different upsampling factors.
This special case is named as 'MDCN-S'.
In order to illustrate the effectiveness of DRB, we provide the following experiments:

(1) In TABLE~\ref{MDCN_S}, we provide PSNR results of MSRN, MDCN-S, and MDCN on 5 benchmark test datasets.
According to this table, we can find: (a) whether MDCN-S or MDCN, the performance is better than MSRN; (b) compared with MDCN-S, the performance of MDCN has been significantly improved.
This proves that the inter-scale correlation can further improve model performance.

\begin{table}
   \centering
   \tiny
   \setlength{\tabcolsep}{1.2mm}
   \renewcommand\arraystretch{1.2}
   \caption{PSNR comparisons of multi-factor models. Best results under different upsampling factors and different training strategies are {\color{red} \textbf{red}}.  Best results under the same upsampling factors with different training strategies are \underline{underline}.}
   \begin{tabular}{|cc|c|c|c|c|c|c|c|c|}
   \hline
                           &      & \multicolumn{2}{c|}{($\times$2)} & \multicolumn{2}{c|}{($\times$3)} & \multicolumn{2}{c|}{($\times$4)} & \multicolumn{2}{c|}{($\times$2,$\times$3,$\times$4)}       \\
                           &   Method   & Urban100      & Manga109      & Urban100      & Manga109      & Urban100      & Manga109      & Urban   & Manga109              \\ \hline \hline
                           & VDSR & 30.77      & 37.22      & *          & *          & *          & *          & 30.80   & 37.33                     \\ \cline{2-10}
                           & MDSR & 32.30      & 38.73      & *          & *          & *          & *          & 32.74   & 38.95                     \\ \cline{2-10}
   \multirow{-3}{*}{x2}    & MDCN & \underline{\color{red} \textbf{32.95}}           & \underline{\color{red} \textbf{39.13}}           & *          & *          & *          & *          &  \color{red} \textbf{32.92} &  \color{red} \textbf{39.09} \\ \hline  \hline
                           & VDSR & *          & *          & 27.14      & 32.01      & *          & *          & 27.25   & 32.20                     \\ \cline{2-10}
                           & MDSR & *          & *          & 28.19      & 33.46      & *          & *          & 28.68   & 34.02                     \\ \cline{2-10}
   \multirow{-3}{*}{x3}    & MDCN & *          & *          &  \color{red} \textbf{28.78}          & \color{red} \textbf{34.11}           & *          & *          & \underline{\color{red} \textbf{28.83}} & \underline{\color{red} \textbf{34.17}} \\ \hline  \hline
                           & VDSR & *          & *          & *          & *          & 25.18      & 28.83      & 25.28   & 28.92                     \\ \cline{2-10}
                           & MDSR & *          & *          & *          & *          & 25.98      & 30.28      & 26.56   & 30.98                     \\ \cline{2-10}
   \multirow{-3}{*}{x4}    & MDCN & *          & *          & *          & *          & \color{red} \textbf{26.45}  & \color{red} \textbf{30.87}           & \underline{\color{red} \textbf{26.69}} & \underline{\color{red} \textbf{31.10}} \\ \hline
   \end{tabular}
   \setlength{\abovecaptionskip}{0.cm}
   \label{multiple}
\end{table}

(2). As described in Section~\ref{hierarchical}, DRB was first proposed in MDSR~\cite{lim2017enhanced}.
Different from MDSR~\cite{lim2017enhanced}, we only apply DRB at the tail of the model.
This means that all feature extraction modules in MDCN are weight sharing, which benefits for inter-scale learning.
In TABLE~\ref{multiple}, we show the quantitative comparisons of multi-factor models, including VDSR~\cite{kim2016accurate}, MDSR~\cite{lim2017enhanced}, and our MDCN.
For a fair comparison, all models are re-trained under the same training dataset and settings.
Among them, ($\times$2), ($\times$3), and ($\times$4) stand for the specifically trained model for one single upsampling factor and ($\times$2,$\times$3,$\times$4) denote the multi-factor model with mix training method.
According to the table, we can find that:
(i). Our MDCN achieves the best results under all upsampling factors whether using the single factor or multi-factor mixed training method, ;
(ii). Most results of multi-factor models (mixed training) are superior to the special training models (single factor training). 
This indicates that fully using the inter-scale correlation between different upsampling factors can improve the model performance;
(iii). At the small upsampling factor ($\times$2), the results of single-factor MDCN are slightly better than the mixed trained MDCN. 
This is because SR images reconstructed with small upsampling factor is a relatively simple task, so that the introduced other scale information will interference it.
Fortunately, this gap is acceptable.
Therefore, we recommend to introduce DRB and mix training strategy to build a general model.

\begin{figure}
   \centering
   \begin{minipage}[c]{0.155\textwidth}
      \includegraphics[width=2.8cm, height=1.6cm]{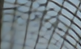} \\
      \includegraphics[width=2.8cm, height=1.6cm]{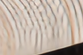} \\
      \includegraphics[width=2.8cm, height=1.6cm]{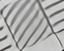}
      \centerline{MSRN~\cite{Li_2018_ECCV}}
   \end{minipage}
   \begin{minipage}[c]{0.155\textwidth}
      \includegraphics[width=2.8cm, height=1.6cm]{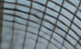} \\
      \includegraphics[width=2.8cm, height=1.6cm]{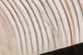} \\ 
      \includegraphics[width=2.8cm, height=1.6cm]{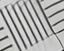}
      \centerline{MDCN (Ours)}
   \end{minipage}
   \begin{minipage}[c]{0.155\textwidth}
      \includegraphics[width=2.8cm, height=1.6cm]{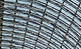} \\ 
      \includegraphics[width=2.8cm, height=1.6cm]{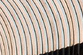} \\ 
      \includegraphics[width=2.8cm, height=1.6cm]{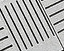}
      \centerline{Ground Truth}
   \end{minipage}
   \caption{Visual comparison between MSRN and MDCN ($\times$4).}
   \label{MSRN}
\end{figure}

\section{Research and Analysis}
\subsection{Compare with MSRN}
Different from MSRN, MDCN introduced MDCB, HFDB, and DRB to build a more efficient and general SR model.
The effectiveness of these modules have been verified in Sec.\ref{Abla}.
Considering the similarity of these models, we provide more detailed comparisons to manifest the effectiveness of MDCN.

(1) In TABLE~\ref{Results}, we provide the PSNR and SSIM results of MSRN and MDCN.
Obviously, MDCN achieves better results.
In addition, we provide some high-frequency detail comparisons between MSRN and MDCN in Fig.~\ref{MSRN}.
We can clearly observe that the high-frequency details of the image reconstructed by MSRN are severely damaged, and even wrong textures and edges are generated.
Contrastly, our MDCN can reconstruct high-quality SR images with accurate and clear details.
% Although there is still a gap compared to the ground truth, this gap is acceptable.
Meanwhile, different from MSRN that specially trained for various upsample factors, we only use one MDCN to perform SR reconstruction task for different factors.

\begin{table}
\tiny
\centering
\setlength{\tabcolsep}{1.2mm}
\renewcommand\arraystretch{1.1}
\caption{Replace the MSRB in the MSRN with MDCB to get the MDCN'. MDCN' achieves better results with fewer parameters.}
\begin{tabular}{|c|c|c|c|c|c|c|}
\hline
Methods    & \multicolumn{3}{c|}{MSRN}                  & \multicolumn{3}{c|}{MDCN'(Ours)}                \\ \hline  \hline
Scale      & x2           & x3           & x4           & x2           & x3           & x4         \\ \hline 
Parameters & 5.92M        & 6.11M        & 6.07M        & {\color{red}\textbf{4.34M}} \textbf{$\downarrow$}    & {\color{red}\textbf{4.52M}} \textbf{$\downarrow$}     & {\color{red}\textbf{4.48M}} \textbf{$\downarrow$}      \\ \hline 
Set5       & 38.07/0.9608 & 34.48/0.9276 & 32.25/0.8958 & \textbf{38.10/0.9608} & \textbf{34.52/0.9278} &  \textbf{32.30/0.8965}     \\ \hline
Set14      & 33.68/0.9184 & 30.40/0.8436 & 28.63/0.7833 & \textbf{33.74/0.9186} & \textbf{30.45/0.8444} &  \textbf{28.68/0.7844}     \\ \hline
BSD100     & 32.22/0.9002 & 29.13/0.8061 & 27.61/0.7377 & \textbf{32.23/0.9003} & \textbf{29.16/0.8067} &  \textbf{27.63/0.7383}     \\ \hline
Urban100   & 32.32/0.9304 & 28.31/0.8560 & 26.20/0.7905 & \textbf{32.34/0.9304} & \textbf{28.39/0.8575} &  \textbf{26.26/0.7918}     \\ \hline
Manga109   & 38.64/0.9771 & 33.56/0.9451 & 30.57/0.9103 & \textbf{38.73/0.9774} & \textbf{33.77/0.9462} &  \textbf{30.68/0.9119}     \\ \hline
Average    & 34.99/0.9374 & 31.18/0.8754 & 29.05/0.8235 & \textbf{35.03/0.9375} &  \textbf{31.26/0.8765} & \textbf{29.11/0.8246}      \\ \hline
\end{tabular}
\label{T-1}
\end{table}

\begin{table}
\tiny
\centering
\setlength{\tabcolsep}{1.2mm}
\renewcommand\arraystretch{1.1}
\caption{Replace the MDCB in the MDCN with MSRB to get the MSRN'. MDCN achieves better results with fewer parameters.}
\begin{tabular}{|c|c|c|c|c|c|c|}
\hline
Methods    & \multicolumn{3}{c|}{MSRN'(x2,x3,x4)}         & \multicolumn{3}{c|}{MDCN (x2,x3,x4, Ours)}                                       \\ \hline  \hline
Scale      & x2           & x3           & x4           & x2           & x3           & x4                                                \\ \hline 
Parameters & \multicolumn{3}{c|}{16.77M}                & \multicolumn{3}{c|}{{\color{red}\textbf{15.62M}} \textbf{$\downarrow$} }        \\ \hline
Set5       & 38.07/0.9608 & 34.51/0.9279 & 32.32/0.8967 & \textbf{38.19/0.9612} & \textbf{34.69/0.9294} & \textbf{32.48/0.8985}           \\ \hline
Set14      & 33.78/0.9192 & 30.45/0.8445 & 28.71/0.7850 & \textbf{33.86/0.9202} & \textbf{30.54/0.8470} & \textbf{28.83/0.7988}           \\ \hline
BSD100     & 32.23/0.9001 & 29.17/0.8070 & 27.66/0.7391 & \textbf{32.32/0.9014} & \textbf{29.26/0.8095} & \textbf{27.74/0.7423}           \\ \hline
Urban100   & 32.43/0.9314 & 28.43/0.8584 & 26.35/0.7946 & \textbf{32.92/0.9355} & \textbf{28.83/0.8662} & \textbf{26.69/0.8049}           \\ \hline
Manga109   & 38.55/0.9779 & 33.72/0.9461 & 30.75/0.9120 & \textbf{39.09/0.9780} & \textbf{34.17/0.9485} & \textbf{31.10/0.9163}           \\ \hline
Average    & 35.01/0.9379 & 31.26/0.8768 & 29.16/0.8255 & \textbf{35.28/0.9393} &  \textbf{31.40/0.8801}  & \textbf{29.37/0.8322}         \\ \hline
\end{tabular}
\label{T-2}
\end{table}

(2) Considering that the parameter amounts of MSRN and MDCN are not equal, we design two experiments to further explore their performance.
To achieve this, we build two models, named MDCN' and MSRN', respectively.
MDCN' is a variant of MSRN, which uses MSRN as the backbone and replaces all MSRBs in the network with MDCB ($C=64$).
In contrast, MSRN' is the variant of MDCN, which uses MDCN as the backbone and replaces all MDCBs in the network with MSRB ($C=128$).
Quantitative comparisons are shown in TABLE~\ref{T-1} and~\ref{T-2}, respectively.
Compared with MSRN, MDCN' achieves better results on all benchmark test datasets with fewer parameters.
This also verifies the feature extraction ability of MDCB is superior to MSRB.
Similarly, MDCN achieves better results than MSRN' with fewer parameters.
This fully proves the effectiveness of MDCN.

In summary, although MSRN is an efficient SR model, the performance can be further improved.
Meanwhile, extensive experiments demonstrate that MDCN can achieve better results than MSRN with fewer parameters.

\subsection{Compare with Other Multi-scale SR Models}
As described in Sec.~\ref{Introduction}, many improved models of MSRN have been proposed such as MSDN~\cite{chang2019multi}, MSFFRN~\cite{qin2020multi}, MWRN~\cite{chang2020accurate}, and MSRCAN~\cite{cao2019single}.
They introduce dense connection, feature fusion, wide-activated, or channel attention mechanisms to improve the performance of MSRN.
However, these models do not pay attention to the core defects of MSRN, so the introduced mechanism does not significantly boost model performance.
To solve this problem, we proposed MDCN.
As shown in TABLE~\ref{others}, we provide the quantitative comparisons of these models.
According to the table, we can clearly see that the performance of MDCN has been greatly improved compared to others.
This further demonstrates the effectiveness of MDCN.

\begin{table}
   \tiny
   \centering
   \setlength{\tabcolsep}{1.8mm}
   \renewcommand\arraystretch{1.5}
   \caption{Comparisons with multi-scale SR models on Urban100.}
   \begin{tabular}{|c|c|c|c|c|c|c|}
   \hline
   Method & MSRN & MSRCAN & MWRN & MSDN & MSFFRN & MDCN (Ours) \\ \hline \hline
   x2    & 32.32/0.9304 & 31.72/0.9242 & 32.46/0.9313 & 32.51/0.9342 & 32.60/0.9326 & \color{red} \textbf{32.92/0.9355} \\ \hline 
   x3    & 28.31/0.8560 & 27.72/0.8397 & 28.40/0.8569 & 28.63/0.8593 & 28.65/0.8619 & \color{red} \textbf{28.83/0.8662} \\ \hline
   x4    & 26.20/0.7905 & 25.75/0.7733 & 26.29/0.7926 & 26.25/0.7931 & 26.47/0.7980 & \color{red} \textbf{26.69/0.8049} \\ \hline
   \end{tabular}
   \label{others}
\end{table}

\begin{figure}
   \centering
   \begin{minipage}[c]{1\textwidth}
   \includegraphics[width=8.6cm,height=3.6cm]{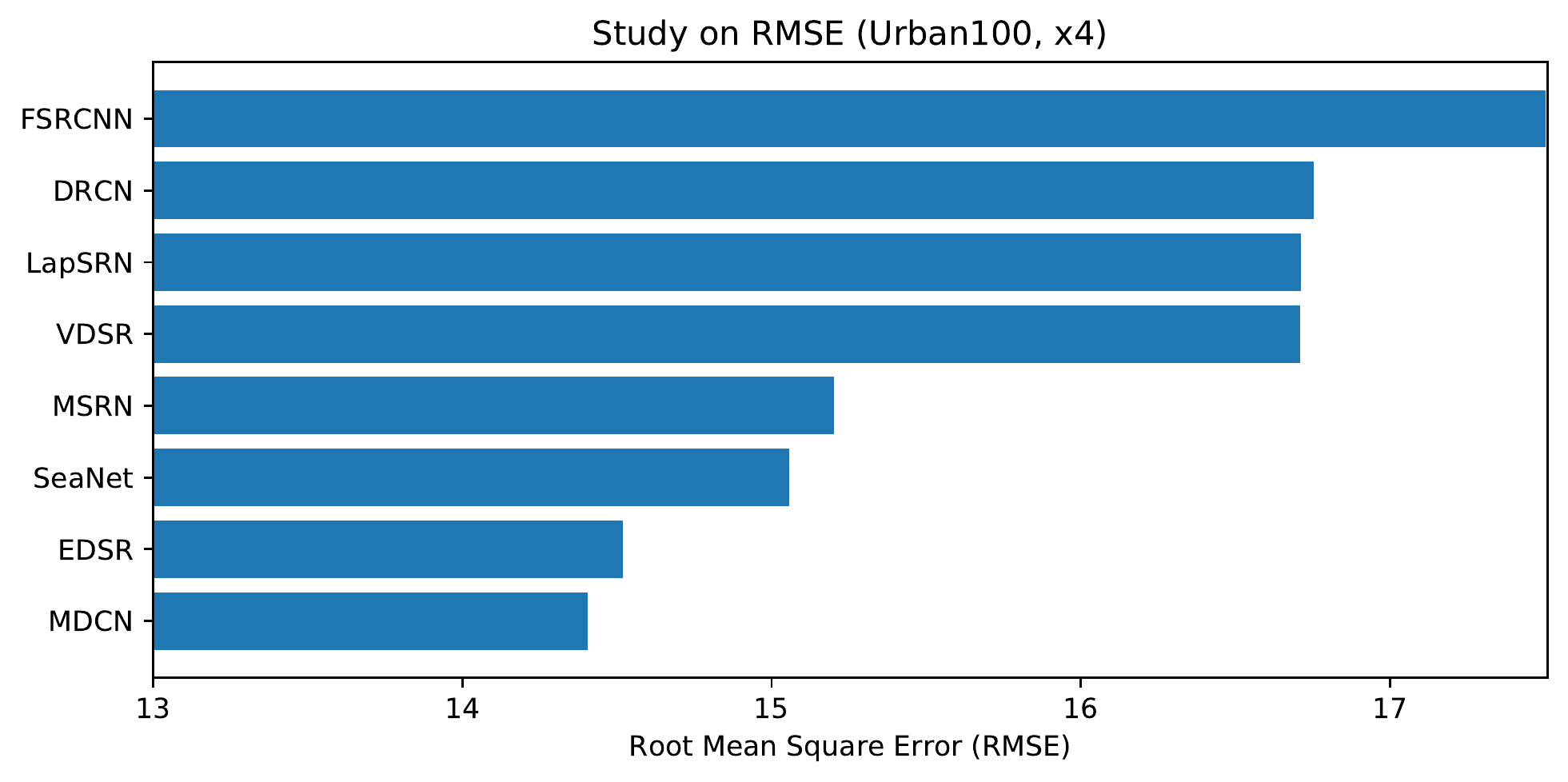}
   \end{minipage}
   \begin{minipage}[c]{1\textwidth}
   \includegraphics[width=8.6cm,height=3.6cm]{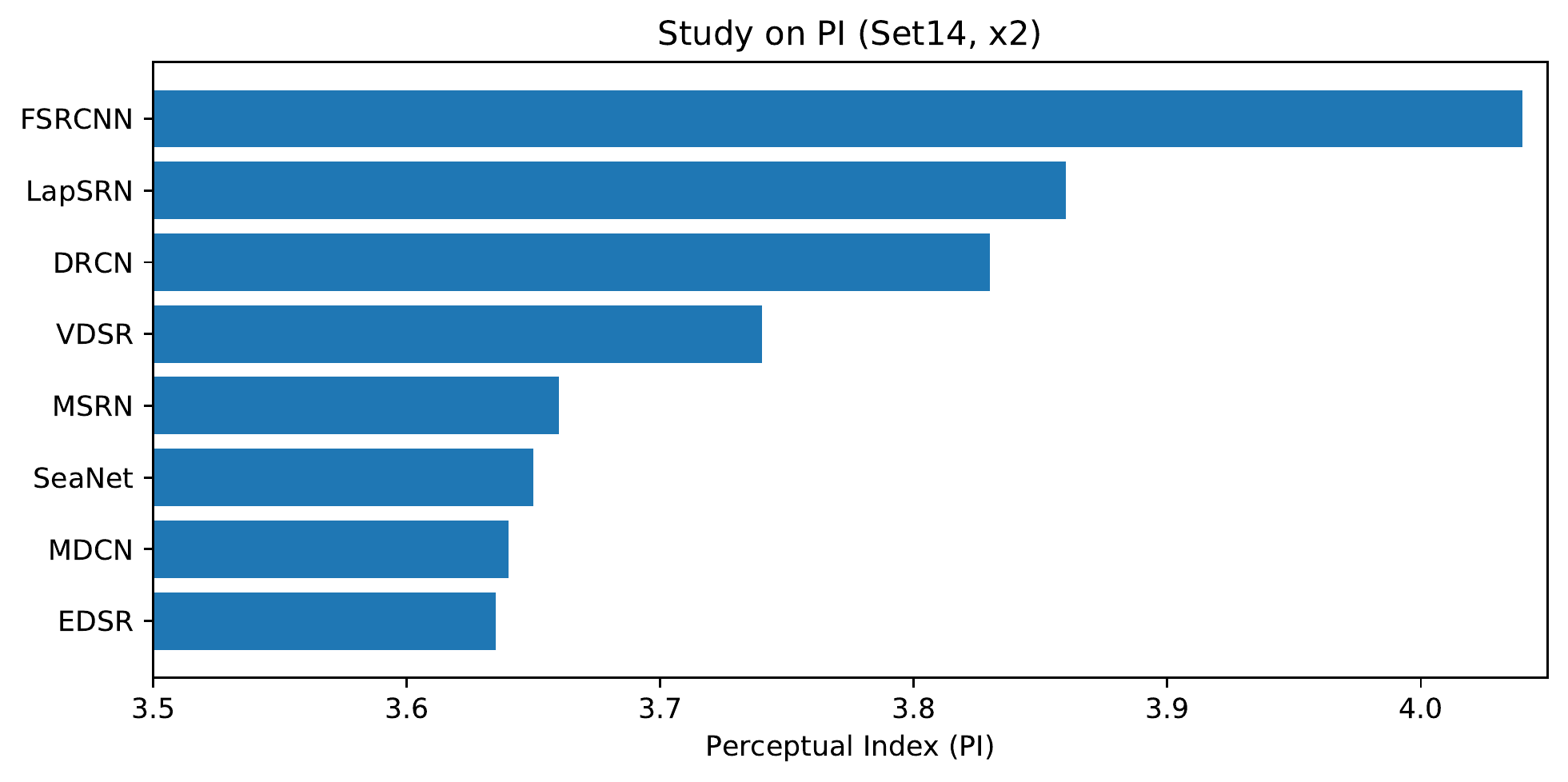}
   \end{minipage}
   \caption{Comparisons of RMSE and PI, respectively. \textbf{A lower index indicates better image quality.}}
   \label{RMSE_PI}
\end{figure}

\subsection{Investigation of Image Naturalness}
In Sec.~\ref{Exp}, we evaluate the performance of MDCN from PSNR, SSIM, and visual effects.
In order to further verify the distribution and perceptual quality of the reconstructed images, we introduce new indicators in this part, including Root Mean Square Error (RMSE) and Perceptual Index~\cite{blau2018perception} (PI).
Among them, RMSE is used to measure the standard deviation of the difference between SR and HR images, and PI is a new criterion that bridges the visual effect with computable index.
Specifically, PI is a perceptual quality index which is judged by the $Ma$ and $NIQE$ scores:
\begin{equation}
   \small
   PI = \frac{1}{2}((10-Ma) + NIQE),
\end{equation}
where $Ma$ is the non-reference measures of Ma’s~\cite{ma2017learning} score and NIQE~\cite{mittal2012making} is a natural image quality evaluator.
Meanwhile, a lower perceptual index ($PI$) represents better perceptual quality.

In Fig.~\ref{RMSE_PI}, we provide the results of RMSE and PI.
According to the top figure, we can clearly see that the RMSE result of MDCN is much lower than other methods.
This means that the SR images reconstructed by MDCN have a closer data distribution to real HR images, in other words, the reconstructed images have higher quality.
For the below one, we can observe that the PI results of EDSR~\cite{lim2017enhanced} and MDCN are smaller than others and the performance of them are very close.
This indicates that the images reconstructed by EDSR and MDCN have better perceptual quality and image naturalness.
All these results fully manifest that our MDCN can achieve competitive performance.

\begin{table}
   \centering
   \scriptsize
   \setlength{\tabcolsep}{1.8mm}
   \renewcommand\arraystretch{1}
   \caption{Image segmentation performance. Due to page limit, only 5 categories are provided and ‘Mean IoU’ denotes the average accuracy of 19 categories.}
   \begin{tabular}{|c|c|c|c|c|c||c|} \hline
   Evaluation & building & traffic light & sky   & car   & truck & Mean IoU \\ \hline \hline
   Bicubic        & 41.38    & 15.08         & 47.82 & 59.03 & 20.59 & 27.17 \\ \hline
   MDCN (Ours)     & 55.59    & 25.08         & 81.34 & 59.63 & 27.71 & 35.61 \\ \hline
   Original Image & 63.15    & 26.26         & 82.16 & 65.08 & 29.97 & 37.19 \\ \hline
   \end{tabular}
   \setlength{\abovecaptionskip}{0.cm}
   \label{seg}
\end{table}

\begin{figure}
   \centering
   \begin{minipage}[c]{0.155\textwidth}
      \includegraphics[width=2.9cm, height=1.65cm]{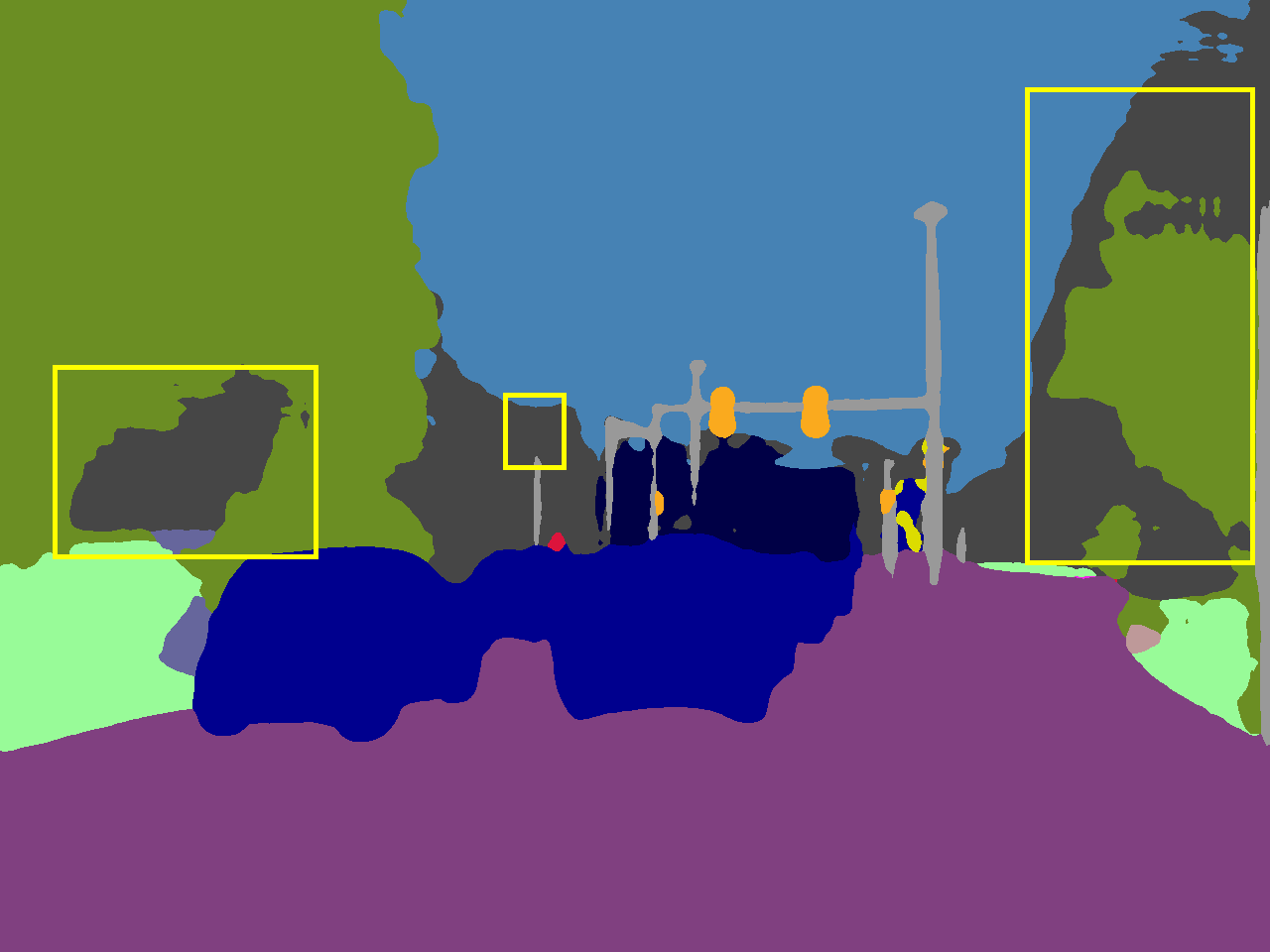}\\
      \includegraphics[width=2.9cm, height=1.65cm]{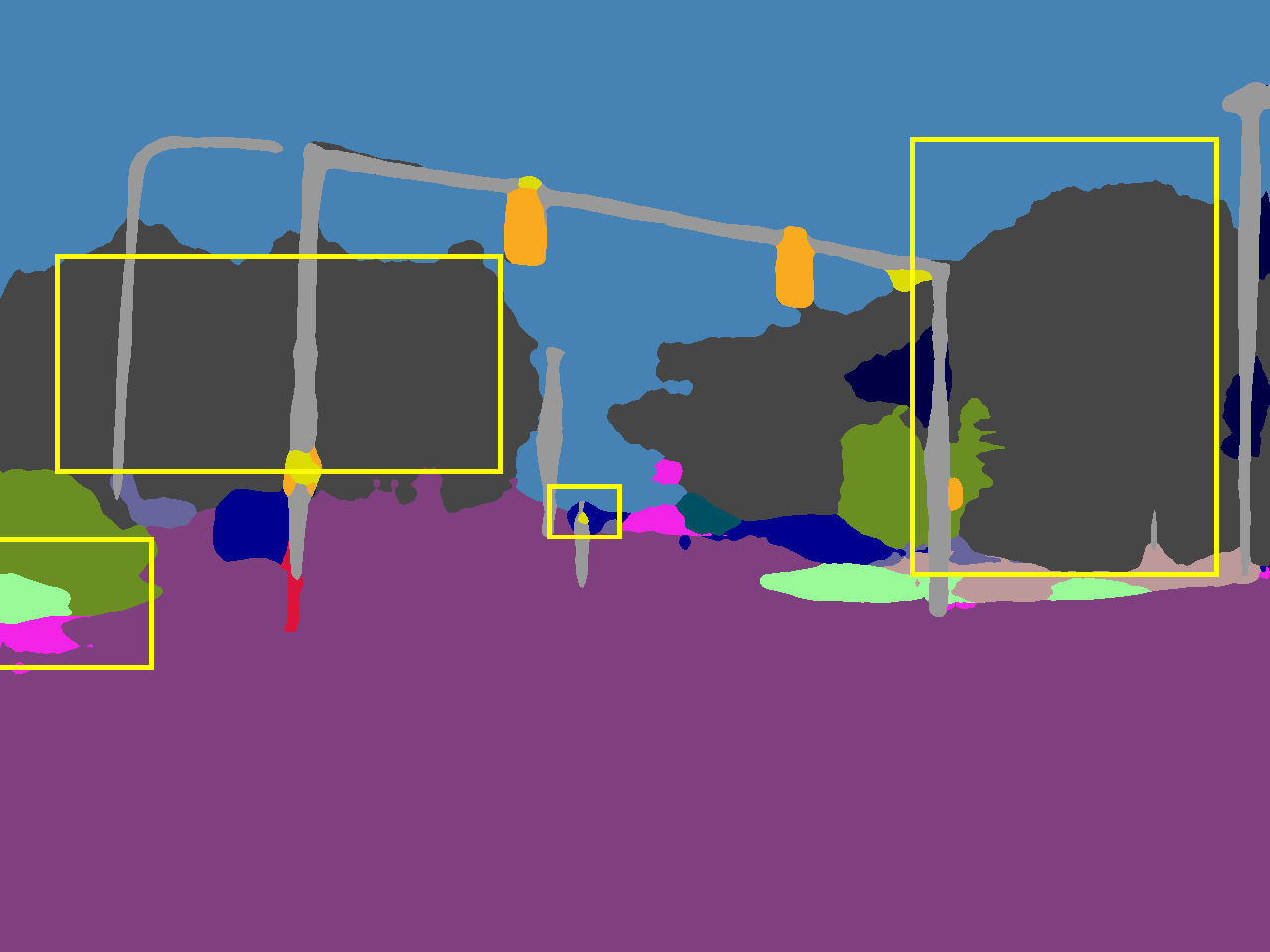}
      \centerline{Bicubic}
   \end{minipage}
   \begin{minipage}[c]{0.155\textwidth}
      \includegraphics[width=2.9cm, height=1.65cm]{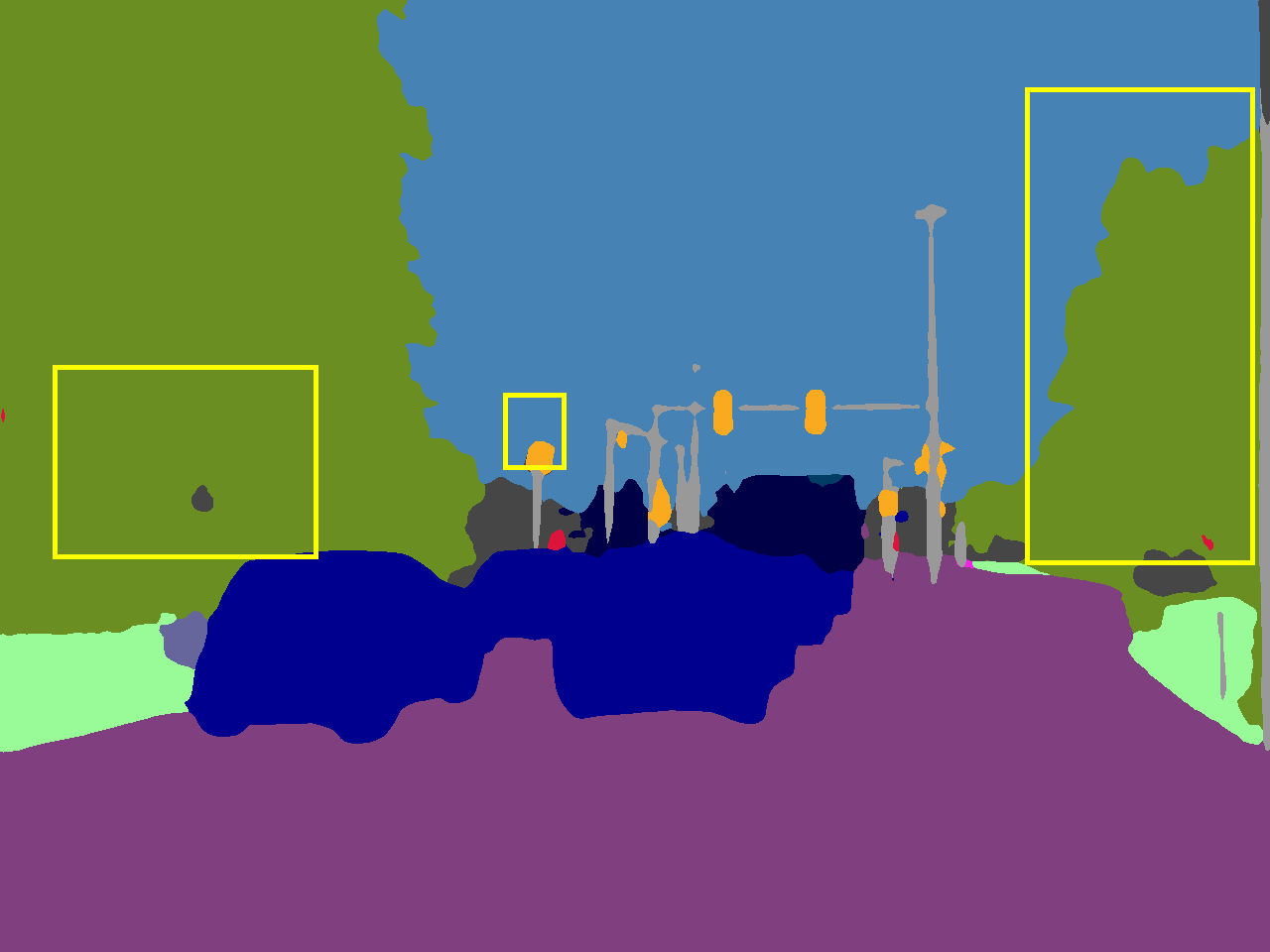}\\
      \includegraphics[width=2.9cm, height=1.65cm]{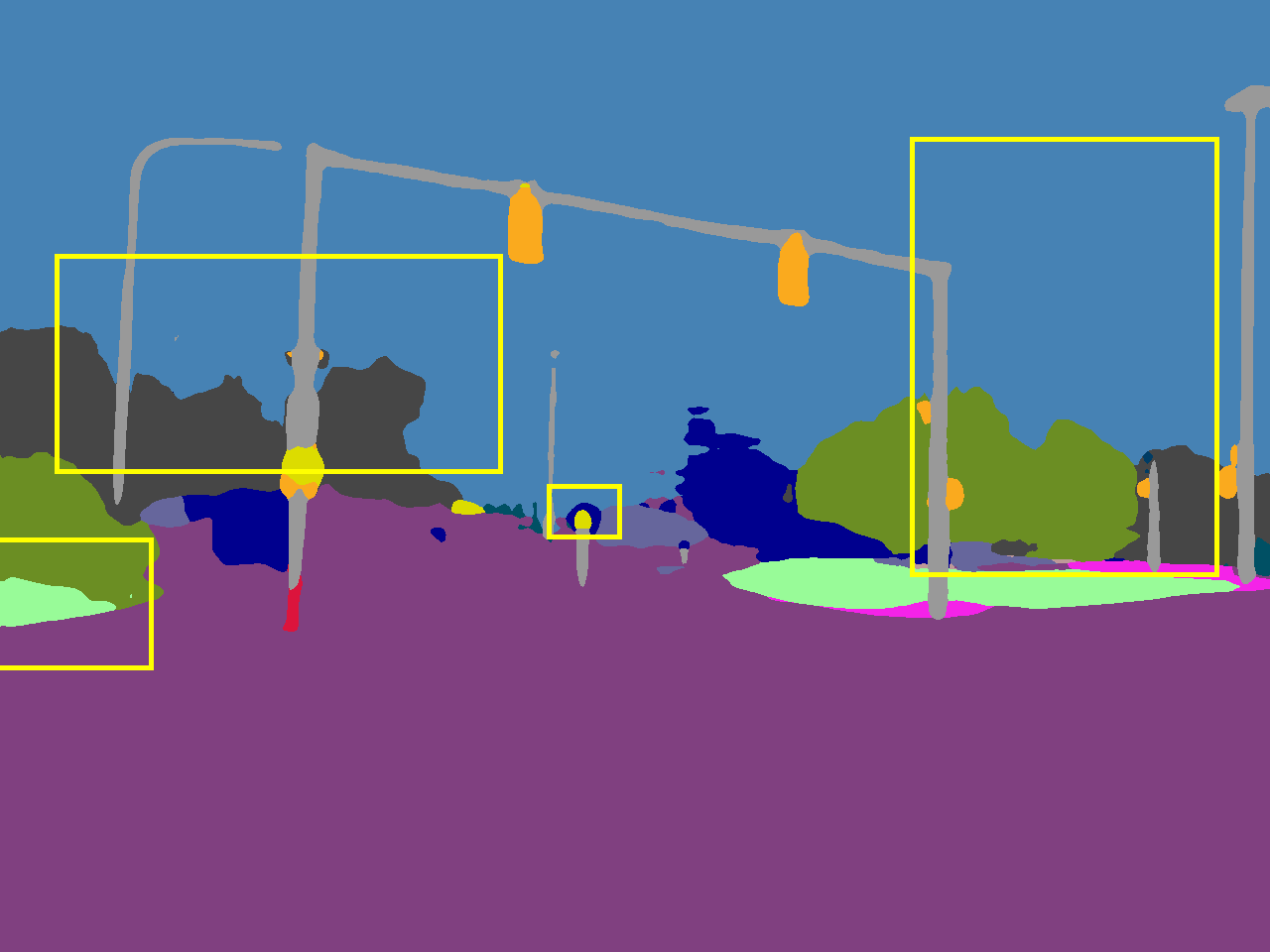}
      \centerline{MDCN (Ours)}
   \end{minipage}
   \begin{minipage}[c]{0.155\textwidth}
      \includegraphics[width=2.9cm, height=1.65cm]{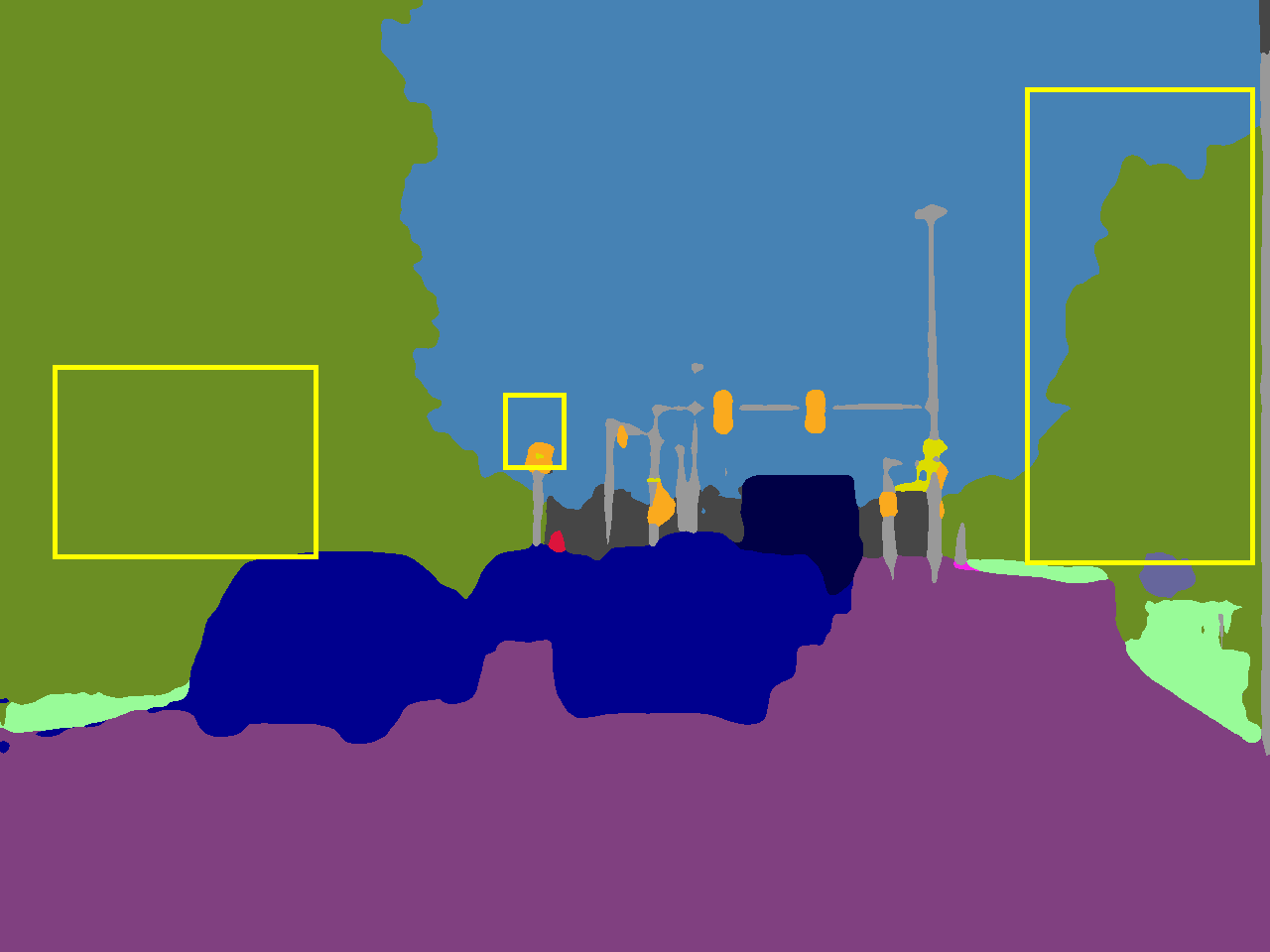}\\
      \includegraphics[width=2.9cm, height=1.65cm]{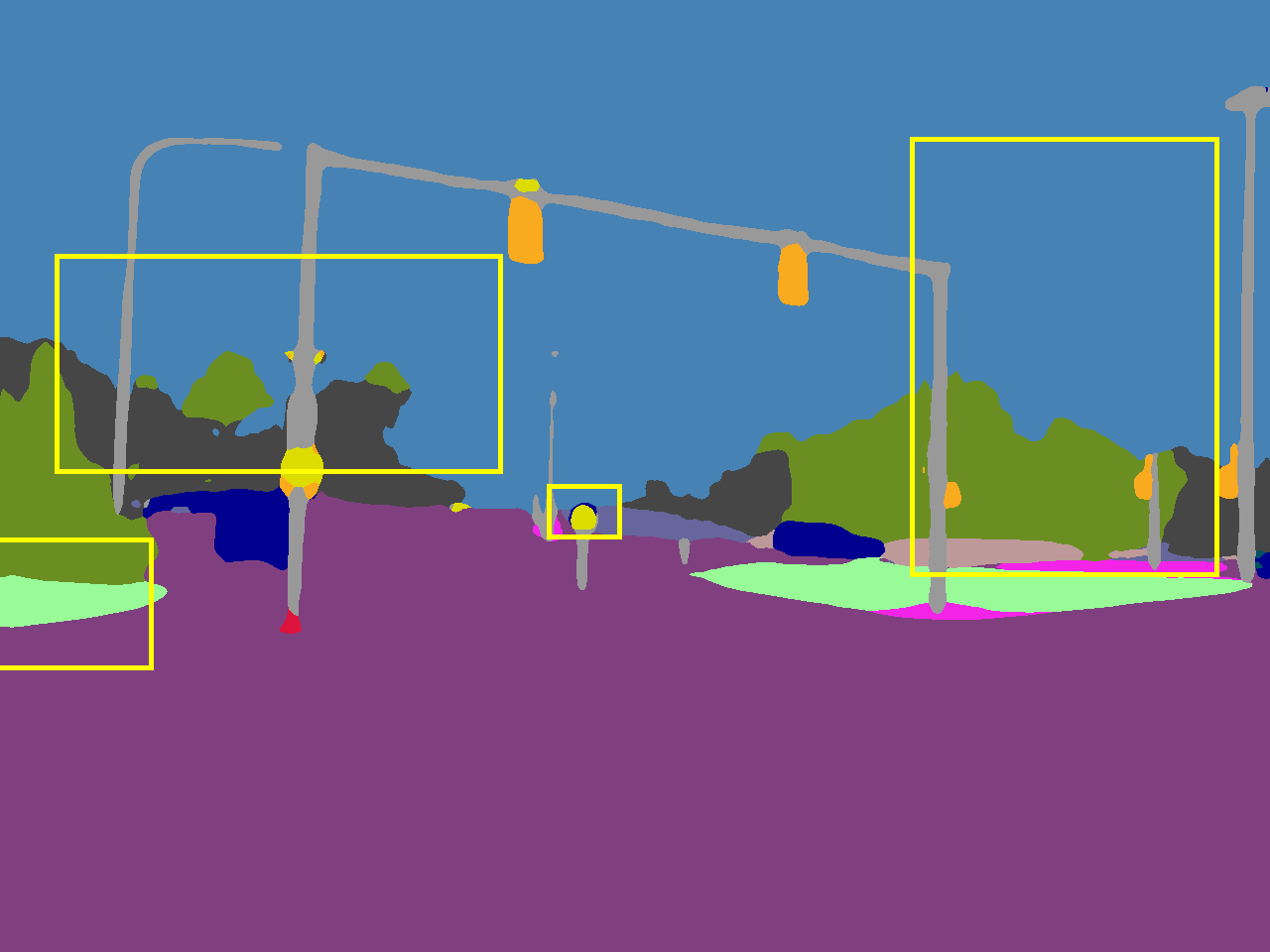}
      \centerline{Original}
   \end{minipage}
   \caption{Comparison of image segmentation results (x4).}
   \label{segmentation}
\end{figure}

\subsection{Exploring on High-level Task}
As we know, the quality of SR images will seriously affect the accuracy of high-level visual tasks such as image segmentation.
In this part, we evaluate the performance of image segmentation to prove it.
We use DRN-38~\cite{Yu2017} as our segmentation model and Cityscapes~\cite{cordts2016cityscapes} (foggy driving dataset) as the test dataset. 
Firstly, we downsample (Bicubic) the original image to obtain the LR input with the factor of $\times$4.
Then, we use Bicubic and our MDCN to reconstruct the corresponding SR image.
Finally, we use DRN-38~\cite{Yu2017} to segment the reconstructed SR image.
In TABLE~\ref{seg}, we provide the IoU results of these methods and we show some segmentation results in Fig.~\ref{segmentation}.
In order to show the accuracy of the segmentation, we mark some areas with green rectangular boxes.
We hope the segmentation result to be consistent with the result marked in "Original", which means the better quality of the reconstructed image. 
According to the figure, we can observe that there are a lot of gray areas in the "Bicubic", which represents the objects are marked incorrectly. 
In contrast, our MDCN obtains better segmentation results. 
This demonstrates that MDCN can reconstruct high-quality images and the quality of SR images will seriously affect the accuracy of high-level visual tasks.
However, it cannot be ignored that there is still a large gap between the image reconstructed by MDCN and the original image.
This means that SISR technology can be further improved.
We will explore the performance of more SISR models on image segmentation or other high-level tasks in future works.
Meanwhile, we aim to combine the feedback from high-level tasks to further boost model performance.

\begin{figure}
   \centering
   \begin{minipage}[c]{1\textwidth}
   \includegraphics[width=8cm,height=3.6cm]{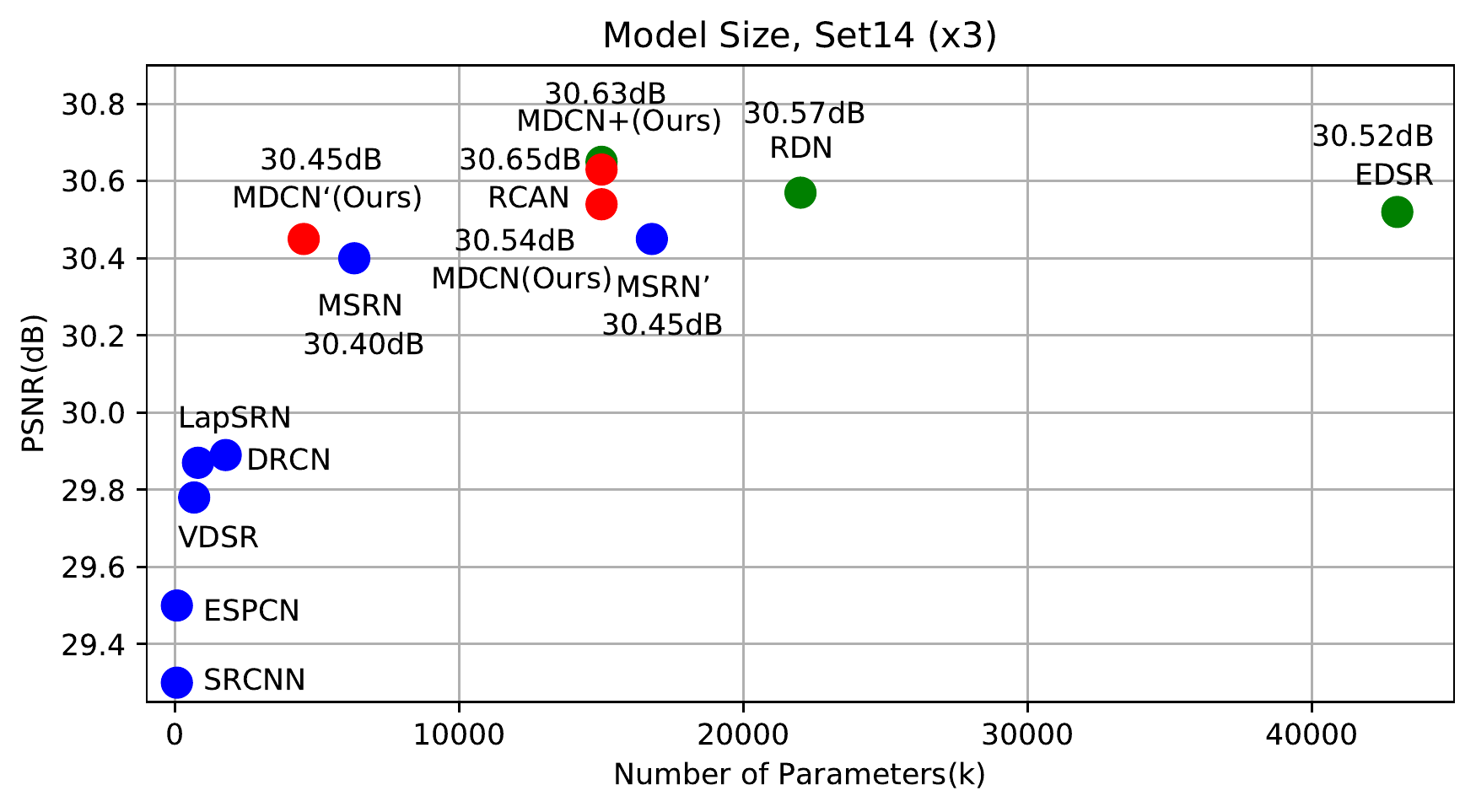}
   \end{minipage}
   \begin{minipage}[c]{1\textwidth}
   \includegraphics[width=8.2cm,height=3.6cm]{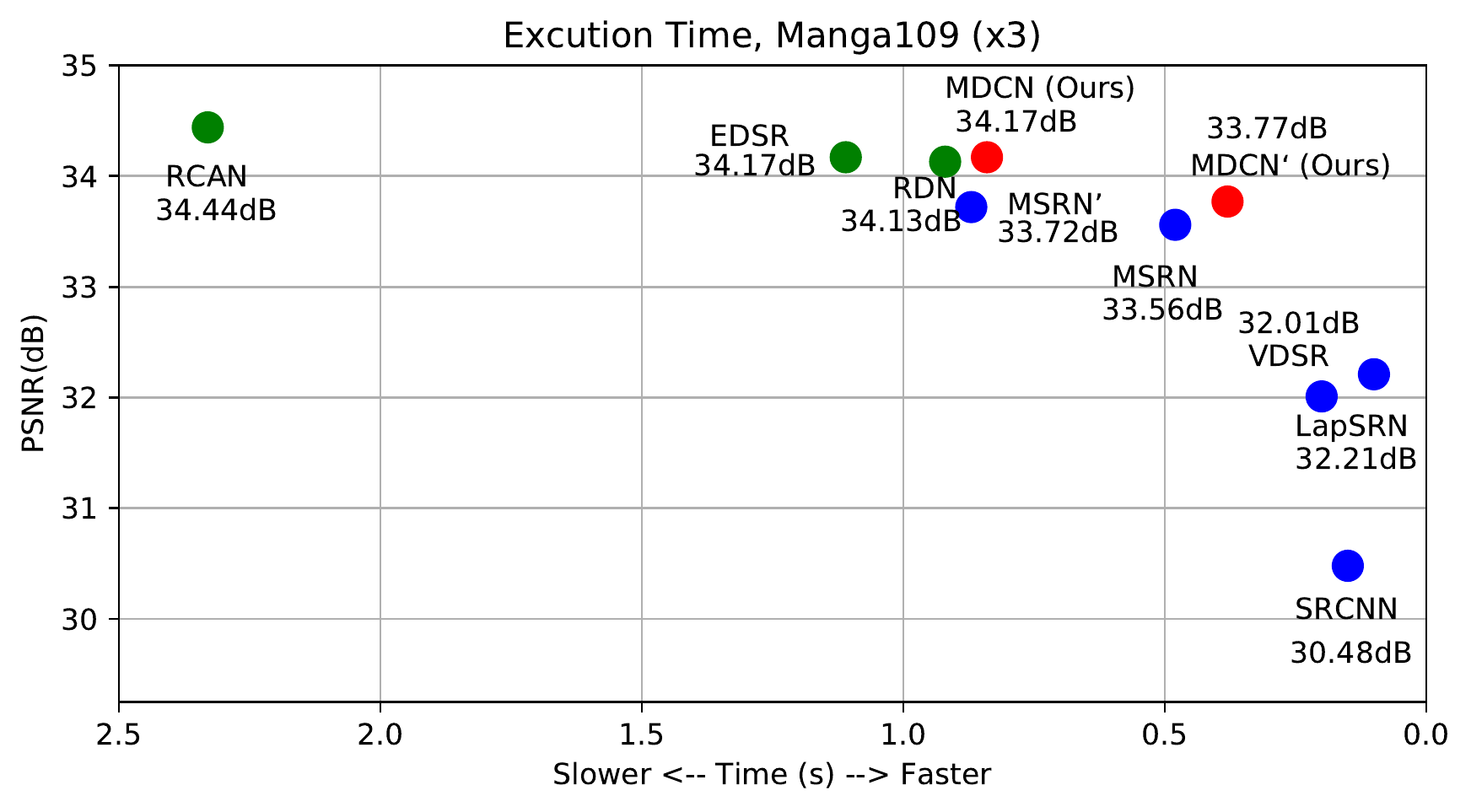}
   \end{minipage}
   \caption{Investigations of the model size and execution time.}
   \label{Size}
\end{figure}

\subsection{Model Size and Execution Time}\label{size}
Increasing the depth of the model is the easiest way to improve model performance.
Therefore, some large size models have been proposed in recent years such as EDSR~\cite{lim2017enhanced} and RDN~\cite{zhang2018residual}.
However, it cannot be ignored that these models are also accompanied by numerous parameters.
It means that these models require more storage space, computing resources, and execution time.
In Fig.~\ref{Size}, we show the comparison of model parameters and execution time.
Among them, red dots represent our MDCN', MDCN, and MDCN+, respectively.
According to the figure, we can draw the following conclusions:
(i) Compared to lightweight SR models (e.g., SRCNN, VDSR, and MSRN), the performance of MDCN is greatly improved; 
(ii) Compared to large models (e.g., EDSR and RDN), MDCN achieves close or better results with fewer parameters and less execution time;
(iii) Compared with RCAN, we can find that RCAN is slightly better than MDCN.
However, it should be noted that the execution time of RCAN is 3 times of MDCN.
Furthermore, MDCN can be suitable for multiple upsampling factors (x2, x3, and x4) without any re-training. 
It means MDCN can save lots of storage space, training time, and computational overhead.
In summary, MDCN achieves a well balance between model performance, model size, and execution time.

\section{Discussion}
Although MDCN is an efficient and general SR model, it still has some limitations:

(1) In this paper, we focus on the reconstruction effect of MDCN on simulated degradation (e.g., Bicubic downsampling) images and verify its effectiveness from multiple indicators (e.g., PSNR, SSIM, RMSE, PI, and visual effect).
It is worth noting that the task of real image super-resolution is more difficult due to the unknown and random degradation modes.
However, this does not mean that our research is meaningless.
Like previous works~\cite{dong2014learning, dong2016accelerating, kim2016accurate, kim2016deeply, shi2016real, lai2017deep, lim2017enhanced, tai2017image, tong2017image, haris2018deep, Li_2018_ECCV, han2018image, zhang2018learning, Hu2018ChannelwiseAS, he2019mrfn, Li2019FilterNetAI, li2019lightweight, Wang2019ResolutionAwareNF, Xie2019FastSS, fang2020soft}, we concentrate on the design of efficient and universal network structure.
Recently, some methods~\cite{Cai2019NTIRE2C, Chen2019OrientationAwareDN, Gao2019MultiscaleDN, kohler2019toward} have been proposed for real image super-resolution.
These models use real-world super-resolution datasets for training, thus the model can achieve real image super-resolution.
This manifests that MDCN can also be implemented to real image super-resolution by fine-tuning on the RealSR dataset.
We will explore the performance of MDCN on real images in the future work.

(2) To make MDCN applicable to multiple upsampling factors without any re-training, we introduce the dynamic reconstruction block (DRB) to learn the inter-scale correlation between different upsampling factors.
However, due to the limitations of sub-pixel convolutional layer, this method cannot directly handle non-integer factors.
In order to remdy this problem, we suggest combining our MDCN with Bicubic to handle any upsampling factors, including non-integer factors.
Specifically, we first utilize MDCN to perform integer magnification, and then use Bicubic to adjust the non-integer part.
For example, for the factor of $\times$3.2, use MDCN to process $\times$3 first, then apply Bicubic to adjust it to the $\times$3.2.
Although the strategy is simple, it is very effective.
At the same time, the performance of this strategy far exceeds the method of using Bicubic to directly enlarge the image.
We also notice that some researchers~\cite{Hu2019MetaSRAM, Wang2020LearningFS} claim that their models can deal with arbitrary upsampling factors.
In the future work, we will further explore the reliability of these strategies and improve our MDCN to handle arbitrary upsampling factors.

\section{Conclusions}
In this paper, we propose a Multi-scale Dense Cross Network (MDCN) for SISR.
MDCN is a robust SR model that can be applied to multiple upsampling factors without any re-training.
Specifically, the proposed multi-scale dense cross blocks (MDCB), hierarchical feature block (HFDB), and dynamic reconstruction block (DRB) together form our MDCN.
Among them, MDCB aims to fully detect and utilize local and multi-scale features, HFDB focuses on maximizing distillation and uses hierarchical features, and the introduced DRB enables the model to learn inter-scale correlations between different upsampling factors.
Extensive evaluations demonstrate that MDCN can achieve competitive results with fewer parameters and less execution time, which achieves an excellent balance between model size and performance.
Furthermore, MDCN can be easily expanded to other low-level computer vision tasks such as image denoising, image dehazing, and image enhancement.
We will further verify the performance of MDCN on other image restoration tasks in future works.

% you can choose not to have a title for an appendix
% if you want by leaving the argument blank

% use section* for acknowledgment
% \section{Acknowledgements}
% This research was supported by the Key Project of the National Natural Science Foundation of China under Grant 61731009, in part by the National Natural Science Foundation of China under Grant 61871185, and in part by the “Chenguang Program” supported by the Shanghai Education Development Foundation and Shanghai Municipal Education Commission under Grant 17CG25. 

% Can use something like this to put references on a page
% by themselves when using endfloat and the captionsoff option.
\ifCLASSOPTIONcaptionsoff
\newpage
\fi

% trigger a \newpage just before the given reference
% number - used to balance the columns on the last page
% adjust value as needed - may need to be readjusted if
% the document is modified later
%\IEEEtriggeratref{8}
% The "triggered" command can be changed if desired:
%\IEEEtriggercmd{\enlargethispage{-5in}}

% references section

% can use a bibliography generated by BibTeX as a .bbl file
% BibTeX documentation can be easily obtained at:
% http://mirror.ctan.org/biblio/bibtex/contrib/doc/
% The IEEEtran BibTeX style support page is at:
% http://www.michaelshell.org/tex/ieeetran/bibtex/
%\bibliographystyle{IEEEtran}
% argument is your BibTeX string definitions and bibliography database(s)
%\bibliography{IEEEabrv,../bib/paper}
%
% <OR> manually copy in the resultant .bbl file
% set second argument of \begin to the number of references
% (used to reserve space for the reference number labels box)

\bibliographystyle{unsrt}
\bibliography{egbib}

\begin{thebibliography}{10}

\bibitem{zhang2006edge}
Lei Zhang and Xiaolin Wu.
\newblock An edge-guided image interpolation algorithm via directional
  filtering and data fusion.
\newblock {\em IEEE Transactions on Image Processing}, 15(8):2226--2238, 2006.

\bibitem{liu2011image}
Xianming Liu, Debin Zhao, Ruiqin Xiong, Siwei Ma, Wen Gao, and Huifang Sun.
\newblock Image interpolation via regularized local linear regression.
\newblock {\em IEEE Transactions on Image Processing}, 20(12):3455--3469, 2011.

\bibitem{dong2013sparse}
Weisheng Dong, Lei Zhang, Rastislav Lukac, and Guangming Shi.
\newblock Sparse representation based image interpolation with nonlocal
  autoregressive modeling.
\newblock {\em IEEE Transactions on Image Processing}, 22(4):1382--1394, 2013.

\bibitem{zhang2018single}
Yunfeng Zhang, Qinglan Fan, Fangxun Bao, Yifang Liu, and Caiming Zhang.
\newblock Single-image super-resolution based on rational fractal
  interpolation.
\newblock {\em IEEE Transactions on Image Processing}, 27(8):3782--3797, 2018.

\bibitem{timofte2013anchored}
Radu Timofte, Vincent De~Smet, and Luc Van~Gool.
\newblock Anchored neighborhood regression for fast example-based
  super-resolution.
\newblock In {\em Proceedings of the IEEE International Conference on Computer
  Vision}, pages 1920--1927, 2013.

\bibitem{timofte2014a+}
Radu Timofte, Vincent De~Smet, and Luc Van~Gool.
\newblock A+: Adjusted anchored neighborhood regression for fast
  super-resolution.
\newblock In {\em ACCV}, pages 111--126, 2014.

\bibitem{yang2012self}
Min-Chun Yang and Yu-Chiang~Frank Wang.
\newblock A self-learning approach to single image super-resolution.
\newblock {\em IEEE Transactions on Multimedia}, 15(3):498--508, 2012.

\bibitem{zhu2014fast}
Zhiliang Zhu, Fangda Guo, Hai Yu, and Chen Chen.
\newblock Fast single image super-resolution via self-example learning and
  sparse representation.
\newblock {\em IEEE Transactions on Multimedia}, 16(8):2178--2190, 2014.

\bibitem{huang2015single}
Jia-Bin Huang, Abhishek Singh, and Narendra Ahuja.
\newblock Single image super-resolution from transformed self-exemplars.
\newblock In {\em Proceedings of the IEEE Conference on Computer Vision and
  Pattern Recognition}, pages 5197--5206, 2015.

\bibitem{dong2014learning}
Chao Dong, Chen~Change Loy, Kaiming He, and Xiaoou Tang.
\newblock Learning a deep convolutional network for image super-resolution.
\newblock In {\em European Conference on Computer Vision}, pages 184--199,
  2014.

\bibitem{kim2016accurate}
Jiwon Kim, Jung Kwon~Lee, and Kyoung Mu~Lee.
\newblock Accurate image super-resolution using very deep convolutional
  networks.
\newblock In {\em Proceedings of the IEEE Conference on Computer Vision and
  Pattern Recognition}, pages 1646--1654, 2016.

\bibitem{dong2016accelerating}
Chao Dong, Chen~Change Loy, and Xiaoou Tang.
\newblock Accelerating the super-resolution convolutional neural network.
\newblock In {\em European Conference on Computer Vision}, pages 391--407,
  2016.

\bibitem{kim2016deeply}
Jiwon Kim, Jung Kwon~Lee, and Kyoung Mu~Lee.
\newblock Deeply-recursive convolutional network for image super-resolution.
\newblock In {\em Proceedings of the IEEE Conference on Computer Vision and
  Pattern Recognition}, pages 1637--1645, 2016.

\bibitem{shi2016real}
Wenzhe Shi, Jose Caballero, Ferenc Husz{\'a}r, Johannes Totz, Andrew~P Aitken,
  Rob Bishop, Daniel Rueckert, and Zehan Wang.
\newblock Real-time single image and video super-resolution using an efficient
  sub-pixel convolutional neural network.
\newblock In {\em Proceedings of the IEEE Conference on Computer Vision and
  Pattern Recognition}, 2016.

\bibitem{lai2017deep}
Wei-Sheng Lai, Jia-Bin Huang, Narendra Ahuja, and Ming-Hsuan Yang.
\newblock Deep laplacian pyramid networks for fast and accurate
  super-resolution.
\newblock In {\em Proceedings of the IEEE Conference on Computer Vision and
  Pattern Recognition}, pages 624--632, 2017.

\bibitem{lim2017enhanced}
Bee Lim, Sanghyun Son, Heewon Kim, Seungjun Nah, and Kyoung~Mu Lee.
\newblock Enhanced deep residual networks for single image super-resolution.
\newblock In {\em Proceedings of the IEEE Conference on Computer Vision and
  Pattern Recognition Workshops}, pages 1132--1140, 2017.

\bibitem{tai2017image}
Ying Tai, Jian Yang, and Xiaoming Liu.
\newblock Image super-resolution via deep recursive residual network.
\newblock In {\em Proceedings of the IEEE Conference on Computer Vision and
  Pattern Recognition}, page~5, 2017.

\bibitem{tong2017image}
Tong Tong, Gen Li, Xiejie Liu, and Qinquan Gao.
\newblock Image super-resolution using dense skip connections.
\newblock In {\em Proceedings of the IEEE International Conference on Computer
  Vision}, 2017.

\bibitem{haris2018deep}
Muhammad Haris, Gregory Shakhnarovich, and Norimichi Ukita.
\newblock Deep back-projection networks for super-resolution.
\newblock In {\em Proceedings of the IEEE Conference on Computer Vision and
  Pattern Recognition}, pages 1664--1673, 2018.

\bibitem{han2018image}
Wei Han, Shiyu Chang, Ding Liu, Mo~Yu, Michael Witbrock, and Thomas~S Huang.
\newblock Image super-resolution via dual-state recurrent networks.
\newblock In {\em Proceedings of the IEEE Conference on Computer Vision and
  Pattern Recognition}, pages 1654--1663, 2018.

\bibitem{zhang2018learning}
Kai Zhang, Wangmeng Zuo, and Lei Zhang.
\newblock Learning a single convolutional super-resolution network for multiple
  degradations.
\newblock In {\em Proceedings of the IEEE Conference on Computer Vision and
  Pattern Recognition}, 2018.

\bibitem{He2019CascadedDN}
Zewei He, Siliang Tang, Jiangxin Yang, Yanlong Cao, Michael~Ying Yang, and
  Yanpeng Cao.
\newblock Cascaded deep networks with multiple receptive fields for infrared
  image super-resolution.
\newblock {\em IEEE Transactions on Circuits and Systems for Video Technology},
  29:2310--2322, 2019.

\bibitem{wu2020multi}
Huapeng Wu, Zhengxia Zou, Jie Gui, Wen-Jun Zeng, Jieping Ye, Jun Zhang, Hongyi
  Liu, and Zhihui Wei.
\newblock Multi-grained attention networks for single image super-resolution.
\newblock {\em IEEE Transactions on Circuits and Systems for Video Technology},
  2020.

\bibitem{zuo2019multi}
Yifan Zuo, Qiang Wu, Yuming Fang, Ping An, Liqin Huang, and Zhifeng Chen.
\newblock Multi-scale frequency reconstruction for guided depth map
  super-resolution via deep residual network.
\newblock {\em IEEE Transactions on Circuits and Systems for Video Technology},
  30(2):297--306, 2019.

\bibitem{Hu2018ChannelwiseAS}
Yanting Hu, Jie Li, Yuanfei Huang, and Xinbo Gao.
\newblock Channel-wise and spatial feature modulation network for single image
  super-resolution.
\newblock {\em IEEE Transactions on Circuits and Systems for Video Technology},
  2018.

\bibitem{Li2019FilterNetAI}
Feng Li, Huihui Bai, and Yao Zhao.
\newblock Filternet: Adaptive information filtering network for accurate and
  fast image super-resolution.
\newblock 2019.

\bibitem{he2019mrfn}
Zewei He, Yanpeng Cao, Lei Du, Baobei Xu, Jiangxin Yang, Yanlong Cao, Siliang
  Tang, and Yueting Zhuang.
\newblock Mrfn: Multi-receptive-field network for fast and accurate single
  image super-resolution.
\newblock {\em IEEE Transactions on Multimedia}, 2019.

\bibitem{li2019lightweight}
Juncheng Li, Yiting Yuan, Kangfu Mei, and Faming Fang.
\newblock Lightweight and accurate recursive fractal network for image
  super-resolution.
\newblock In {\em ICCVW}, 2019.

\bibitem{Wang2019ResolutionAwareNF}
Yifan Wang, Lijun Wang, Hongyu Wang, and Peihua Li.
\newblock Resolution-aware network for image super-resolution.
\newblock {\em IEEE Transactions on Circuits and Systems for Video Technology},
  29:1259--1269, 2019.

\bibitem{Xie2019FastSS}
Chao Xie, Weili Zeng, and Xiaobo Lu.
\newblock Fast single-image super-resolution via deep network with component
  learning.
\newblock {\em IEEE Transactions on Circuits and Systems for Video Technology},
  29:3473--3486, 2019.

\bibitem{fang2020soft}
Faming Fang, Juncheng Li, and Tieyong Zeng.
\newblock Soft-edge assisted network for single image super-resolution.
\newblock {\em IEEE Transactions on Image Processing}, 29:4656--4668, 2020.

\bibitem{Li_2018_ECCV}
Juncheng Li, Faming Fang, Kangfu Mei, and Guixu Zhang.
\newblock Multi-scale residual network for image super-resolution.
\newblock In {\em European Conference on Computer Vision}, 2018.

\bibitem{Zhang_2018_ECCV}
Yulun Zhang, Kunpeng Li, Kai Li, Lichen Wang, Bineng Zhong, and Yun Fu.
\newblock Image super-resolution using very deep residual channel attention
  networks.
\newblock In {\em European Conference on Computer Vision}, 2018.

\bibitem{yang2019deep}
Wenming Yang, Xuechen Zhang, Yapeng Tian, Wei Wang, Jing-Hao Xue, and Qingmin
  Liao.
\newblock Deep learning for single image super-resolution: A brief review.
\newblock {\em IEEE Transactions on Multimedia}, 2019.

\bibitem{Wang2020DeepLF}
Zhihao Wang, Jian Chen, and Steven C.~H. Hoi.
\newblock Deep learning for image super-resolution: A survey.
\newblock {\em IEEE Transactions on Pattern Analysis and Machine Intelligence},
  2020.

\bibitem{cao2019single}
Feilong Cao and Huan Liu.
\newblock Single image super-resolution via multi-scale residual channel
  attention network.
\newblock {\em Neurocomputing}, 358:424--436, 2019.

\bibitem{sang2019multi}
Yu~Sang, Jinguang Sun, Simiao Wang, Yanfei Peng, Xinjun Zhang, and Zhiyang
  Yang.
\newblock Multi-scale information distillation network for image super
  resolution in nsct domain.
\newblock In {\em International Conference on Neural Information Processing},
  pages 50--59, 2019.

\bibitem{qin2020multi}
Jinghui Qin, Yongjie Huang, and Wushao Wen.
\newblock Multi-scale feature fusion residual network for single image
  super-resolution.
\newblock {\em Neurocomputing}, 379:334--342, 2020.

\bibitem{li2019multi}
Zhuangzi Li, Shanshan Li, Naiguang Zhang, Lei Wang, and Ziyu Xue.
\newblock Multi-scale invertible network for image super-resolution.
\newblock In {\em Proceedings of the ACM Multimedia Asia}, pages 1--6. 2019.

\bibitem{soh2020lightweight}
Jae~Woong Soh and Nam~Ik Cho.
\newblock Lightweight single image super-resolution with multi-scale spatial
  attention networks.
\newblock {\em IEEE Access}, 8:35383--35391, 2020.

\bibitem{wang2020remote}
Xinying Wang, Yingdan Wu, Yang Ming, and Hui Lv.
\newblock Remote sensing imagery super resolution based on adaptive multi-scale
  feature fusion network.
\newblock {\em Sensors}, 20(4):1142, 2020.

\bibitem{zhang2020scene}
Shu Zhang, Qiangqiang Yuan, Jie Li, Jing Sun, and Xuguo Zhang.
\newblock Scene-adaptive remote sensing image super-resolution using a
  multiscale attention network.
\newblock {\em IEEE Transactions on Geoscience and Remote Sensing}, 2020.

\bibitem{zhang2019multi}
Wei-Liang Zhang, Qin-Yan Zhang, Ji-Jiang Yang, and Qing Wang.
\newblock Multi-scale network with the deeper and wider residual block for mri
  motion artifact correction.
\newblock In {\em IEEE Annual Computer Software and Applications Conference},
  volume~2, pages 405--410, 2019.

\bibitem{qi2020pulmonary}
Yongjun Qi, Junhua Gu, Weixun Li, Zepei Tian, Yajuan Zhang, and Juanping Geng.
\newblock Pulmonary nodule image super-resolution using multi-scale deep
  residual channel attention network with joint optimization.
\newblock {\em The Journal of Supercomputing}, 76(2):1005--1019, 2020.

\bibitem{sang2019contourlet}
Yu~Sang, Jinguang Sun, Simiao Wang, Xiangfu Meng, and Heng Qi.
\newblock Contourlet transform based seismic signal denoising via multi-scale
  information distillation network.
\newblock In {\em Pacific Rim International Conference on Artificial
  Intelligence}, pages 660--672, 2019.

\bibitem{szegedy2015going}
Christian Szegedy, Wei Liu, Yangqing Jia, Pierre Sermanet, Scott Reed, Dragomir
  Anguelov, Dumitru Erhan, Vincent Vanhoucke, and Andrew Rabinovich.
\newblock Going deeper with convolutions.
\newblock In {\em Proceedings of the IEEE Conference on Computer Vision and
  Pattern Recognition}, pages 1--9, 2015.

\bibitem{szegedy2016rethinking}
Christian Szegedy, Vincent Vanhoucke, Sergey Ioffe, Jon Shlens, and Zbigniew
  Wojna.
\newblock Rethinking the inception architecture for computer vision.
\newblock In {\em Proceedings of the IEEE Conference on Computer Vision and
  Pattern Recognition}, pages 2818--2826, 2016.

\bibitem{chollet2017xception}
Fran{\c{c}}ois Chollet.
\newblock Xception: Deep learning with depthwise separable convolutions.
\newblock In {\em Proceedings of the IEEE Conference on Computer Vision and
  Pattern Recognition}, pages 1251--1258, 2017.

\bibitem{lai2017fast}
Wei-Sheng Lai, Jia-Bin Huang, Narendra Ahuja, and Ming-Hsuan Yang.
\newblock Deep laplacian pyramid networks for fast and accurate
  super-resolution.
\newblock In {\em Proceedings of the IEEE Conference on Computer Vision and
  Pattern Recognition}, pages 5835--5843, 2017.

\bibitem{zhang2018residual}
Yulun Zhang, Yapeng Tian, Yu~Kong, Bineng Zhong, and Yun Fu.
\newblock Residual dense network for image super-resolution.
\newblock In {\em Proceedings of the IEEE Conference on Computer Vision and
  Pattern Recognition}, 2018.

\bibitem{Russell2019FeatureBasedIP}
Mosin Russell, Ju~Jia Zou, Gu~Fang, and Weidong Cai.
\newblock Feature-based image patch classification for moving shadow detection.
\newblock {\em IEEE Transactions on Circuits and Systems for Video Technology},
  29:2652--2666, 2019.

\bibitem{Han2020DoubleRR}
Na~Ra Han, Jigang Wu, Xiaozhao Fang, Wai~Keung Wong, Yong Xu, Jian Yang, and
  Xuelong Li.
\newblock Double relaxed regression for image classification.
\newblock {\em IEEE Transactions on Circuits and Systems for Video Technology},
  30:307--319, 2020.

\bibitem{Wang2019HierarchicalIS}
Huiqun Wang, Di~Huang, Kui Jia, and Yunhong Wang.
\newblock Hierarchical image segmentation ensemble for objectness in rgb-d
  images.
\newblock {\em IEEE Transactions on Circuits and Systems for Video Technology},
  29:93--103, 2019.

\bibitem{Zhou2018ComputationAM}
Yiren Zhou, Thanh-Toan Do, Haitian Zheng, Ngai-Man Cheung, and Lu~Fang.
\newblock Computation and memory efficient image segmentation.
\newblock {\em IEEE Transactions on Circuits and Systems for Video Technology},
  28:46--61, 2018.

\bibitem{Li2020HeadNetAE}
Wei Li, Hongliang Li, Qingbo Wu, Fanman Meng, Linfeng Xu, and King~Ngi Ngan.
\newblock Headnet: An end-to-end adaptive relational network for head
  detection.
\newblock {\em IEEE Transactions on Circuits and Systems for Video Technology},
  30:482--494, 2020.

\bibitem{Sommer2019ComprehensiveAO}
Lars~Wilko Sommer, Tobias Schuchert, and J{\"u}rgen Beyerer.
\newblock Comprehensive analysis of deep learning-based vehicle detection in
  aerial images.
\newblock {\em IEEE Transactions on Circuits and Systems for Video Technology},
  29:2733--2747, 2019.

\bibitem{he2016deep}
Kaiming He, Xiangyu Zhang, Shaoqing Ren, and Jian Sun.
\newblock Deep residual learning for image recognition.
\newblock In {\em Proceedings of the IEEE Conference on Computer Vision and
  Pattern Recognition}, pages 770--778, 2016.

\bibitem{huang2017densely}
Gao Huang, Zhuang Liu, Laurens Van Der~Maaten, and Kilian~Q Weinberger.
\newblock Densely connected convolutional networks.
\newblock In {\em Proceedings of the IEEE Conference on Computer Vision and
  Pattern Recognition}, volume~1, page~3, 2017.

\bibitem{hu2018squeeze}
Jie Hu, Li~Shen, and Gang Sun.
\newblock Squeeze-and-excitation networks.
\newblock In {\em Proceedings of the IEEE Conference on Computer Vision and
  Pattern Recognition}, pages 7132--7141, 2018.

\bibitem{mei2018effective}
Kangfu Mei, Aiwen Jiang, Juncheng Li, Jihua Ye, and Mingwen Wang.
\newblock An effective single-image super-resolution model using
  squeeze-and-excitation networks.
\newblock In {\em International Conference on Neural Information Processing},
  pages 542--553, 2018.

\bibitem{agustsson2017ntire}
Eirikur Agustsson and Radu Timofte.
\newblock Ntire 2017 challenge on single image super-resolution: Dataset and
  study.
\newblock In {\em Proceedings of the IEEE Conference on Computer Vision and
  Pattern RecognitionW}, pages 1110--1121, 2017.

\bibitem{bevilacqua2012}
Marco Bevilacqua, Aline Roumy, Christine Guillemot, and Marie~Line
  Alberi-Morel.
\newblock Low-complexity single-image super-resolution based on nonnegative
  neighbor embedding.
\newblock In {\em The British Machine Vision Conference}, 2012.

\bibitem{zeyde2010single}
Roman Zeyde, Michael Elad, and Matan Protter.
\newblock On single image scale-up using sparse-representations.
\newblock In {\em International Conference on Curves and Surfaces}, pages
  711--730, 2010.

\bibitem{martin2001database}
David Martin, Charless Fowlkes, Doron Tal, and Jitendra Malik.
\newblock A database of human segmented natural images and its application to
  evaluating segmentation algorithms and measuring ecological statistics.
\newblock In {\em Proceedings of the IEEE International Conference on Computer
  Vision}, pages 416--423, 2001.

\bibitem{matsui2017sketch}
Yusuke Matsui, Kota Ito, Yuji Aramaki, Azuma Fujimoto, Toru Ogawa, Toshihiko
  Yamasaki, and Kiyoharu Aizawa.
\newblock Sketch-based manga retrieval using manga109 dataset.
\newblock {\em Multimedia Tools and Applications}, 76(20):21811--21838, 2017.

\bibitem{arbelaez2011}
Pablo Arbelaez, Michael Maire, Charless Fowlkes, and Jitendra Malik.
\newblock Contour detection and hierarchical image segmentation.
\newblock {\em IEEE Transactions on Pattern Analysis and Machine Intelligence},
  33(5):898--916, 2011.

\bibitem{huang2015singl}
Jia-Bin Huang, Abhishek Singh, and Narendra Ahuja.
\newblock Single image super-resolution from transformed self-exemplars.
\newblock In {\em Proceedings of the IEEE Conference on Computer Vision and
  Pattern Recognition}, pages 5197--5206, 2015.

\bibitem{timofte2016seven}
Radu Timofte, Rasmus Rothe, and Luc Van~Gool.
\newblock Seven ways to improve example-based single image super resolution.
\newblock In {\em Proceedings of the IEEE Conference on Computer Vision and
  Pattern Recognition}, pages 1865--1873, 2016.

\bibitem{wang2004image}
Zhou Wang, Alan~C Bovik, Hamid~R Sheikh, and Eero~P Simoncelli.
\newblock Image quality assessment: from error visibility to structural
  similarity.
\newblock {\em IEEE Transactions on Image Processing}, 13(4):600--612, 2004.

\bibitem{chang2019multi}
Chia-Yang Chang and Shao-Yi Chien.
\newblock Multi-scale dense network for single-image super-resolution.
\newblock In {\em IEEE International Conference on Acoustics, Speech and Signal
  Processing}, pages 1742--1746, 2019.

\bibitem{chang2020accurate}
Kan Chang, Minghong Li, Pak Lun~Kevin Ding, and Baoxin Li.
\newblock Accurate single image super-resolution using multi-path
  wide-activated residual network.
\newblock {\em Signal Processing}, page 107567, 2020.

\bibitem{blau2018perception}
Yochai Blau and Tomer Michaeli.
\newblock The perception-distortion tradeoff.
\newblock In {\em Proceedings of the IEEE Conference on Computer Vision and
  Pattern Recognition}, pages 6228--6237, 2018.

\bibitem{ma2017learning}
Chao Ma, Chih-Yuan Yang, Xiaokang Yang, and Ming-Hsuan Yang.
\newblock Learning a no-reference quality metric for single-image
  super-resolution.
\newblock {\em Computer Vision and Image Understanding}, 158:1--16, 2017.

\bibitem{mittal2012making}
Anish Mittal, Rajiv Soundararajan, and Alan~C Bovik.
\newblock Making a “completely blind” image quality analyzer.
\newblock {\em IEEE Signal Processing Letters}, 20(3):209--212, 2012.

\bibitem{Yu2017}
Fisher Yu, Vladlen Koltun, and Thomas Funkhouser.
\newblock Dilated residual networks.
\newblock In {\em Proceedings of the IEEE Conference on Computer Vision and
  Pattern Recognition}, 2017.

\bibitem{cordts2016cityscapes}
Marius Cordts, Mohamed Omran, Sebastian Ramos, Timo Rehfeld, Markus Enzweiler,
  Rodrigo Benenson, Uwe Franke, Stefan Roth, and Bernt Schiele.
\newblock The cityscapes dataset for semantic urban scene understanding.
\newblock In {\em Proceedings of the IEEE Conference on Computer Vision and
  Pattern Recognition}, pages 3213--3223, 2016.

\bibitem{Cai2019NTIRE2C}
Jianrui Cai, Shuhang Gu, Radu Timofte, Lei Zhang, and et~al.
\newblock Ntire 2019 challenge on real image super-resolution: Methods and
  results.
\newblock In {\em Proceedings of the IEEE Conference on Computer Vision and
  Pattern Recognition Workshops}, 2019.

\bibitem{Chen2019OrientationAwareDN}
Du~Chen, Zewei He, Anshun Sun, Jiangxin Yang, Yanlong Cao, Yanpeng Cao, Siliang
  Tang, and Michael~Ying Yang.
\newblock Orientation-aware deep neural network for real image
  super-resolution.
\newblock In {\em Proceedings of the IEEE Conference on Computer Vision and
  Pattern Recognition Workshops}, 2019.

\bibitem{Gao2019MultiscaleDN}
Shangqi Gao and Xiahai Zhuang.
\newblock Multi-scale deep neural networks for real image super-resolution.
\newblock In {\em Proceedings of the IEEE Conference on Computer Vision and
  Pattern Recognition Workshops}, 2019.

\bibitem{kohler2019toward}
Thomas K{\"o}hler, Michel B{\"a}tz, Farzad Naderi, Andr{\'e} Kaup, Andreas
  Maier, and Christian Riess.
\newblock Toward bridging the simulated-to-real gap: Benchmarking
  super-resolution on real data.
\newblock {\em IEEE Transactions on Pattern Analysis and Machine Intelligence},
  2019.

\bibitem{Hu2019MetaSRAM}
Xuecai Hu, Haoyuan Mu, Xiangyu Zhang, Zilei Wang, Tieniu Tan, and Jian Sun.
\newblock Meta-sr: A magnification-arbitrary network for super-resolution.
\newblock {\em Proceedings of the IEEE Conference on Computer Vision and
  Pattern Recognition}, pages 1575--1584, 2019.

\bibitem{Wang2020LearningFS}
Longguang Wang, Yingqian Wang, Zaiping Lin, Jungang Yang, Wei An, and Yulan
  Guo.
\newblock Learning for scale-arbitrary super-resolution from scale-specific
  networks.
\newblock 2020.

\end{thebibliography}

% \begin{thebibliography}{1}

% \bibitem{IEEEhowto:kopka}
% H.~Kopka and P.~W. Daly, \emph{A Guide to \LaTeX}, 3rd~ed.\hskip 1em plus 0.5em minus 0.4em\relax Harlow, England: Addison-Wesley, 1999.

% \end{thebibliography}

% biography section
%
% If you have an EPS/PDF photo (graphicx package needed) extra braces are
% needed around the contents of the optional argument to biography to prevent
% the LaTeX parser from getting confused when it sees the complicated
% \includegraphics command within an optional argument. (You could create
% your own custom macro containing the \includegraphics command to make things
% simpler here.)
%\begin{IEEEbiography}[{\includegraphics[width=1in,height=1.25in,clip,keepaspectratio]{mshell}}]{Michael Shell}
% or if you just want to reserve a space for a photo:

% \begin{IEEEbiography}{ Shell}
% Biography text here.
% \end{IEEEbiography}

% if you will not have a photo at all:
% \begin{IEEEbiography}{John Doe}
% Biography text here.
% \end{IEEEbiography}

% insert where needed to balance the two columns on the last page with
% biographies
%\newpage

% \begin{IEEEbiography}{Jane Doe}
% Biography text here.
% \end{IEEEbiography}

% You can push biographies down or up by placing
% a \vfill before or after them. The appropriate
% use of \vfill depends on what kind of text is
% on the last page and whether or not the columns
% are being equalized.

%\vfill

% Can be used to pull up biographies so that the bottom of the last one
% is flush with the other column.
%\enlargethispage{-5in}

% that's all folks
\end{document}